\documentclass[amsmath,amssymb,aip,jcp,reprint,longbibliography]{revtex4-1}

\usepackage{graphicx}% Include figure files
\usepackage{dcolumn}% Align table columns on decimal point
\usepackage{bm}% bold math
\usepackage{color}
\usepackage[]{placeins}

\newcommand{\bra}[1]{\ensuremath{\left\langle #1\right|}}
\newcommand{\ket}[1]{\ensuremath{\left|#1\right\rangle}}
\newcommand{\bigbraket}[2]{\left\langle {#1} \mathrel{\left | {\vphantom {#1 #2}} \right. \kern-\nulldelimiterspace} {#2} \right\rangle}

\newcommand{\thh}[1]{\ensuremath{#1^\textrm{th}}}

\newcommand{\kkk} {\boldsymbol{k} }
\newcommand{\ki} {\ensuremath{\boldsymbol{k}_\mathrm{I} }}
\newcommand{\kii} {\ensuremath{\boldsymbol{k}_\mathrm{II} }}

\newcommand{\ev}{\ensuremath{\mathrm{eV}}}

\begin{document}
\preprint{Multidimensional X-Ray Spectroscopy}

\title{Multidimensional X-Ray Spectroscopy of Valence and Core Excitations in Cysteine}

\author{Jason D. Biggs}
\affiliation{Dept. of Chemistry, University of California, 450 Rowland Hall, Irvine,
California 92697, USA}

\author{Yu Zhang}
\affiliation{Dept. of Chemistry, University of California, 450 Rowland Hall, Irvine,
California 92697, USA}

\author{Daniel Healion}
\affiliation{Dept. of Chemistry, University of California, 450 Rowland Hall, Irvine, California 92697, USA}

\author{Shaul Mukamel}

\email{smukamel@uci.edu}

\affiliation{Dept. of Chemistry, University of California, 450 Rowland Hall, Irvine,
California 92697, USA}
\date{\today}
\begin{abstract}
Several  nonlinear spectroscopy experiments which employ broadband x-ray pulses to probe the coupling between localized core and delocalized valence excitation are simulated for the amino acid cysteine at the K-edges of oxygen and nitrogen and the K and L-edges of sulfur.
We focus on two dimensional (2D) and 3D signals generated by two- and three-pulse stimulated x-ray Raman spectroscopy (SXRS) with frequency-dispersed probe.
We show how the four-pulse x-ray signals $\ki=-\kkk_1+\kkk_2+\kkk_3$ and  $\kii=\kkk_1-\kkk_2+\kkk_3$ can give new 3D insight into the SXRS signals.  The coupling between valence- and core-excited states can be visualized in three dimensional plots, revealing the origin of the polarizability that controls the simpler pump-probe SXRS signals.
\end{abstract}

\maketitle

\section{Introduction}
Ultrashort x-ray pulses may be used to probe valence excitations prepared by a stimulated x-ray Raman process.\cite{healion_simulation_2011,biggs_two-dimensional_2012,zhang:194306,mukamel_mutidimensional_2013} In this technique, an inner-shell (core) electron is excited to the valence band by an ultrashort laser pulse.  Then, after the molecule evolves in this highly excited state for a time limited by the pulse duration, the same pulse stimulates the emission of a photon and the core-hole is filled.  The final state can be any valence-excited state (or the ground state) lying within the pulse bandwidth.  Because the core to valence transition frequency is characteristic to the element from which the core electron is being excited, this technique can be spatially selective.  For example, in a molecule containing a single nitrogen atom, Raman excitation by a pulse tuned to the nitrogen core-edge ($\sim 400\, \mathrm{eV}$) excites only those valence-excited states perturbed by the presence of a core-hole localized at the
nitrogen atom.

Stimulated Raman is one component of the pump-probe signal, which measures the response by
exciting the system with a short \emph{pump} pulse, waiting for a period
$\tau$ and then measuring the transmission of the \emph{probe} pulse.
The central frequencies of the two pulses control which dynamics are
probed during the interval, $\tau$.
Shorter pulses and more precise time measurements have allowed
dynamics at shorter length and time-scales to be detected as the technology has improved.

Time-resolved spectroscopy\cite{Mukamel1995} has been widely employed in nuclear magnetic
resonance (NMR) to detect transitions between nuclear spin levels split
by an external magnetic field, and is sensitive to dynamics on the
$\mu$s timescale.\cite{ernst_principles_1990} Spin states have long lifetimes and are easily manipulated by external fields. Linear techniques were extended to the nonlinear regime, where multiple pulses are used to prepare and probe
nonstationary states.\cite{ernst_principles_1990}
One landmark was the spin echo
technique,\cite{hahn_free_1953,hahn_spin_1950} where a simple two pulse sequence is capable of
discriminating between
homogeneous and inhomogeneous line broadening, due to fast and slow frequency fluctuations, respectively.

The photon echo,\cite{abella_photon_1966} an extension of
the spin echo technique into the optical regime,
has been applied to molecular electronic transitions\cite{Aartsma1976520,cooper_intermolecular_1980}
and later to vibrational systems such
as the amide bond stretch in
proteins\cite{hamm_structure_1998,asplund_two-dimensional_2000} and
the hydrogen bond network in water.\cite{cowan_ultrafast_2005}  Pulses
with central frequencies tuned to optical transitions have been used
to investigate photosynthetic
complexes,\cite{doi:10.1146/annurev.physchem.040808.090259,Scholes_2010}
and semiconductors .\cite{cundiff_optical_2009}

The development of x-ray free electron
lasers (XFELs) \cite{ullrich_free-electron_2012,Glownia:10} and
higher harmonic generation\cite{gallmann_attosecond_2012} sources may enable
pump-probe and photon-echo experiments at x-ray frequencies.
Zholents and Penn have proposed a method to generate attosecond pulse pairs, each member of which
is independently tunable through the soft x-ray region.\cite{zholents_obtaining_2010}
Their simulations predict that the technique will be able to produce $\sim 250 \, \textrm{as}$
pulses (bandwidth 8.5 eV) with center frequencies at the nitrogen and oxygen K-edges, like those used
the simulations presented in Section \ref{sec:sim}.
Guimar\~{a}es and Gel'mukhanov simulated
infrared (IR) pump/x-ray probe in a model diatomic molecule.\cite{guimaraes_pump-probe_2006}  They proposed that
core-hole delocalization in homonuclear diatomics can be inferred from this signal.   In a model system with temporally overlapping optical pump and x-ray probe pulses, it was shown that when the
probe pulse duration is shorter than one cycle of the pump field, the signal is highly dependent on the
absolute phase of the optical pump.\cite{guimaraes_two-color_2004}

Optical pump/ x-ray probe
experiments have already been reported in atoms.\cite{goulielmakis_real-time_2010}
 Recent work on electron and hole
mobilities following ultrafast photoionization has also stimulated
great interest in nonlinear x-ray experiments.\cite{kuleff_tracing_2007,sansone_electron_2012,peng_probe_2012} Currently,
most studies use the collection of molecular ion or photoelectrons to
measure interactions with a probing beam, rather than the change in
intensity of the probe.\cite{tzallas_extreme-ultraviolet_2011}  Other possibilities which incorporate noisy x-ray pulses have
been proposed.\cite{meyer_noisy_2012}

Two-pulse pump-probe experiments, in which the signal is defined as the total integrated intensity of the probe with the pump minus that without the pump, were simulated in Refs. \onlinecite{healion_simulation_2011,zhang:194306}. These signals were integrated over all frequencies, and recorded versus the interpulse delay.  Interaction with the pump pulse will create core-excitations through direct absorption and valence excitations via a Raman process. By selecting delay times which are longer than the lifetime of the core-excited state ($\leq 10 \, \mathrm{fs}$), only valence contributions to the signal survive.  This integrated two-pulse stimulated x-ray Raman spectroscopy (I2P-SXRS) signal is a direct probe of the valence response to a localized core hole which is switched on and then off again during the x-ray pulse.  A three-pulse (3P) extension I3P-SXRS, proposed in Ref. \onlinecite{biggs_two-dimensional_2012}, has two interpulse delays, and thus two dimensions.
Information on the core-excited intermediate state, the higher-lying scattering state, can only be obtained indirectly in I2P-SXRS and I3P-SXRS.  Increased information is available from these pump-probe signals by frequency-dispersing the probe pulse rather than recording only its integrated intensity, giving an extra dimension to the signals and allowing a look \emph{inside} the probe polarizability.

Frequency-domain x-ray Raman spectroscopy, known as resonant x-ray inelastic scattering (RIXS), is a well-established technique.  Here excitation is achieved using a monochromatic beam, and the spontaneous emission spectrum is recorded as a function of excitation frequency. The simplest time-domain x-ray Raman technique, I2P-SXRS, uses two pulses, the pump and probe, and measures the pump-induced change in the transmitted intensity of the probe, integrated over all frequencies. The signal is recorded versus the delay time between pump and probe.  Due to the vastly shorter lifetimes of core holes compared with valence excitations, we can solely focus on valence excitations by examining the I2P-SXRS signal for delay times longer than the core lifetime ($\sim 10 \, \mathrm{fs}$).  I2P-SXRS has additional capabilities that RIXS does not.  It is possible to use a two-color setup where the pump and probe are  resonant with different core transitions, revealing those valence excitations perturbed by both core holes.
Further, the signal arising from the isotropic polarizability can be selected by using a magic angle configuration of pulse polarizations.

Here we add an extra dimension to these techniques by measuring the pump-induced change in probe transmission, as a function of delay time \emph{and} the probe frequency.  D2P-SXRS shows which core-excited states are coupled to which valence-excited states by peering inside the probe-pulse effective polarizability.

\section{Nonlinear X-ray Signals}
\begin{figure}[htbp]
  \includegraphics[width = 8.5 cm] {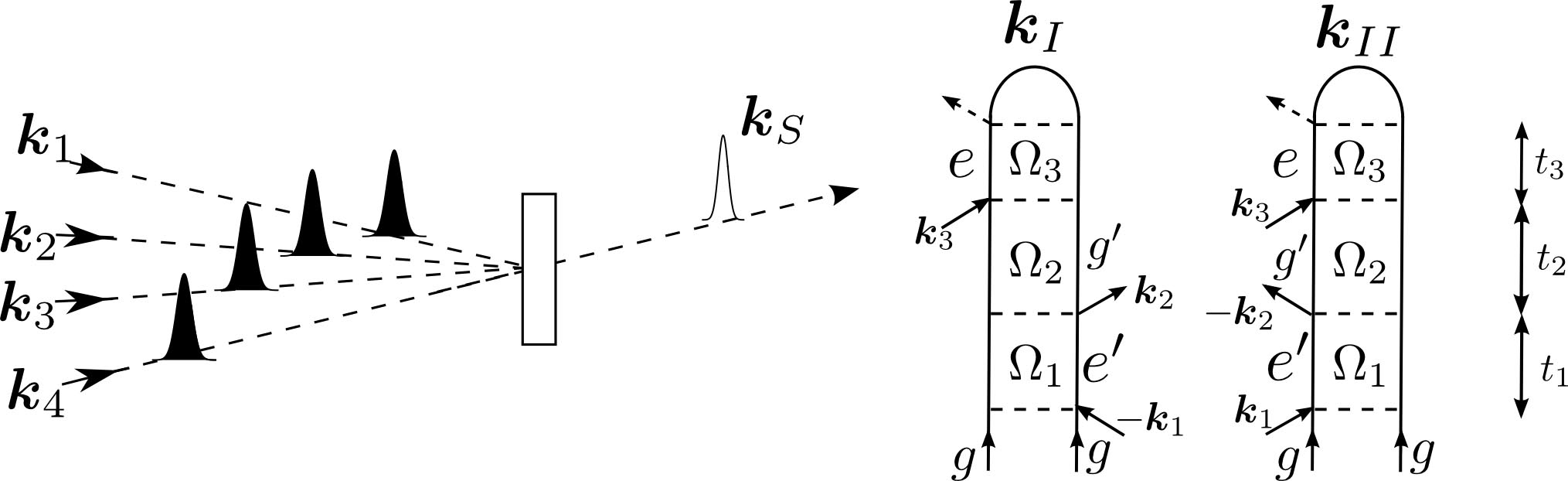}
\caption{\label{fig:XPE-diags}(Left) Pulse sequence for the four-wave mixing setup.  (Right) The contributing loop diagrams for the $\ki$ and $\kii$ signals.}
 \end{figure}

Here we outline the signals considered in this work. In four-wave mixing, the molecule interacts with four short pulses, and the signal is given by
the change in transmission of the fourth pulse induced by the other
pulses. The signals generated in the directions $\ki=-\kkk_1+\kkk_2+\kkk_3$ and $\kii=\kkk_1-\kkk_2+\kkk_3$ are represented by
the loop diagrams in Fig. \ref{fig:XPE-diags}. In these diagrams, the left and right branches of the loop represent the time evolution of the ket and bra respectively, and real time flows from bottom to top.\cite{biggs_coherent_2011,Mukamel2010223}
Interactions with the field are represented by arrows into (absorption) and out of (emission) the diagram.
The signal is recorded versus the
three delay times $t_1$, $t_2$, and $t_3$, and subsequently Fourier
transformed to give the 3D signal.  Since the system is in a core-excited coherence during the $t_1$ and $t_3$ periods, the conjugate frequencies $\Omega_1$ and $\Omega_3$ will show resonances in the hundreds of eVs.  During the $t_2$ period, known as the waiting time or population time, the system is in a valence-excited coherence and therefore $\Omega_2$ resonances occur at valence frequencies (between 5 and 12 eV for the cysteine model used here).  Expressions for the $\ki$ and $\kii$ signals are given in Appendix \ref{app:4wm}.

\begin{figure}[htbp]
  \includegraphics[width = 1.75 in] {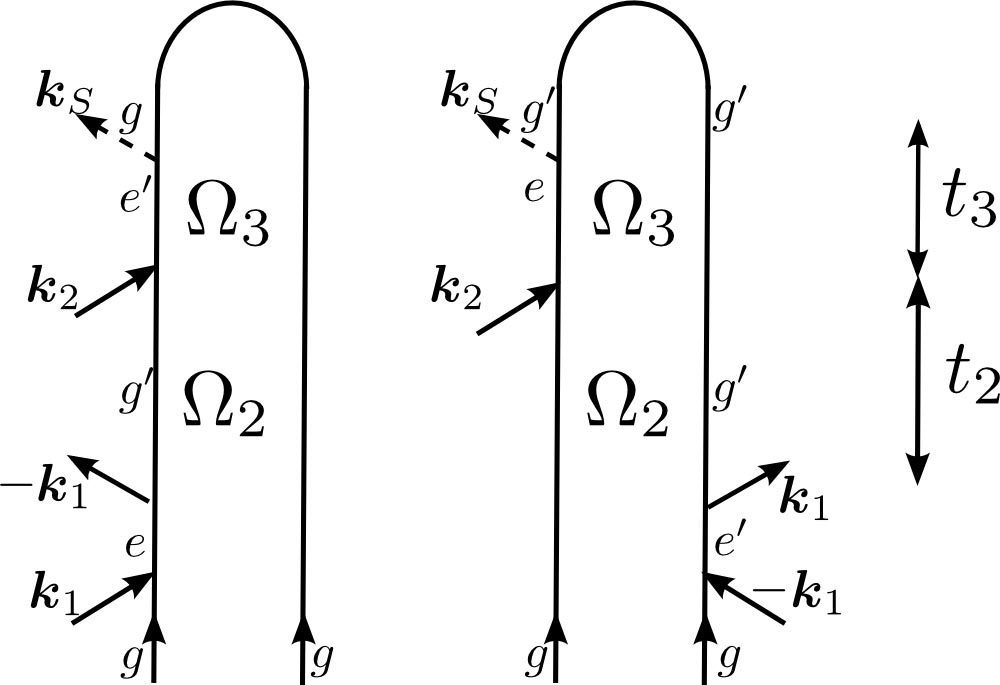}
\caption{\label{fig:1}Loop diagrams for the two-pulse SXRS signal.}
 \end{figure}

In a two-pulse SXRS pump-probe experiment (Fig. \ref{fig:1}), the system interacts with two pulses with a single time delay.  The integrated intensity change of the probe pulse recorded versus the time delay $t_2$ gives the one-dimensional I2P-SXRS signal.  A two-dimensional signal, herein called D2P-SXRS, is obtained by further recording the dispersed spectrum of the probe.  The diagrams in Fig. \ref{fig:1} are the same diagrams contributing the $\ki$ and $\kii$ signals, but with $t_1=0$.  Signals are displayed versus $\Omega_2$, the Fourier conjugate of the single time delay. $\Omega_2$ peaks will show valence excitations.  Expressions for the I2P-SXRS and D2P-SXRS signals are given in Appendix \ref{app:2pSXRS}.

\begin{figure}[htbp]
  \includegraphics[width = 3.5 in] {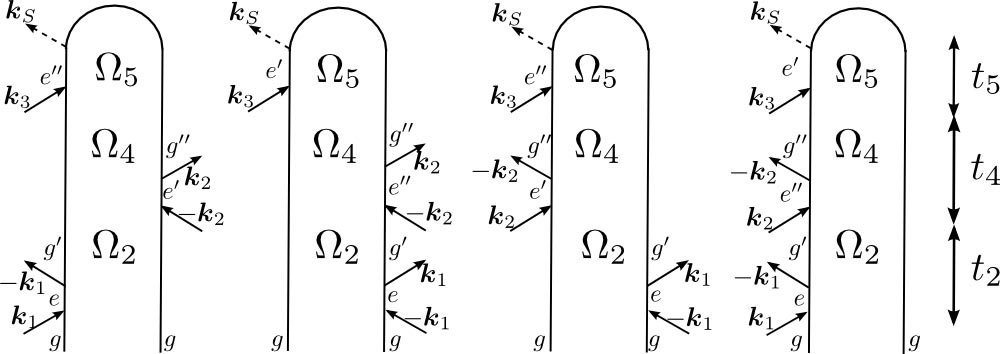}
\caption{\label{fig:2}Loop diagrams for the three-pulse 2D-SXRS signal.  $\Omega_2$ and $\Omega_4$ are the Fourier conjugates of the two delay times, and $\Omega_5$ is the monochrometer frequency.  Peaks along $\Omega_2$ and $\Omega_4$ occur at valence frequencies, and those along $\Omega_5$ at core-excitation frequencies.}
 \end{figure}

The three-pulse SXRS signals, both dispersed and integrated, are represented by the diagrams in Fig. \ref{fig:2}.  The I3P-SXRS signal is defined as the integrated transmitted intensity of the probe pulse with two prior pumps, minus that without. The two inter-pulse delays, $t_2$ and $t_4$, constitute the dimensions of the signal.  A third dimension can be recovered by frequency dispersing the third pulse. Expressions for the I3P-SXRS and D3P-SXRS signals are given in Appendix \ref{app:3pSXRS}.

\section{Simulations}\label{sec:sim}
\subsection{Setup}
We have simulated the spectroscopic techniques shown in Figs. \ref{fig:XPE-diags}--\ref{fig:2} for the amino acid cysteine.
This small sulfur-containing molecule provides an important structural function
by connecting different regions of proteins through disulfide bonds.  It has been implicated in biological charge transfer in the respiratory
complex I.\cite{hayashi_electron_2010}  The optimized geometry of cysteine was obtained with the Gaussian09
package\cite{G09} at the B3LYP\cite{Becke93,SDCF94}/6-311G** level of
theory. All restricted excitation windown (REW) TDDFT calculations and
transition dipole calculations for the core excited states were
performed with a locally modified version of NWChem
code \cite{NWChem,LKKG12} at the CAM-B3LYP\cite{YTH04}/6-311G** level
of theory, and with the Tamm-Dancoff approximation.\cite{HHG99b} For
more computational details please see Ref. \onlinecite{zhang:194306}.

\begin{figure}[htbp]
  \includegraphics[width = 8.5 cm] {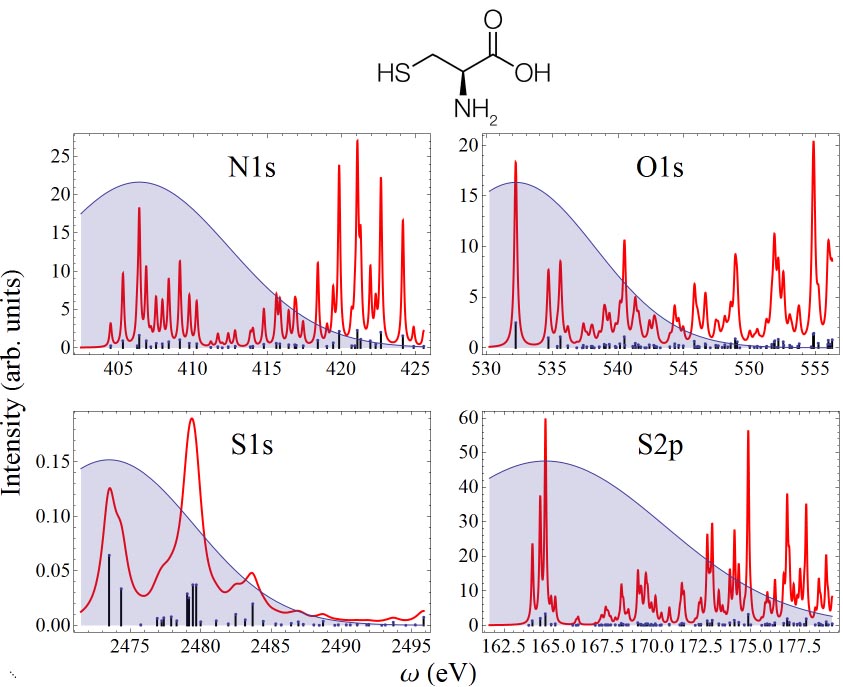}
\caption{\label{fig:xanes}Simulated x-ray absorption (XANES) spectra for cysteine at the nitrogen, oxygen, and sulfur K-edges, and the sulfur L2-edge.  Stick spectra are shown in black, and simulated spectra with constant Lorentzian linewidths are shown in red.    The following linewidths were used: $\Gamma_{\mathrm{N1s}}= 0.09 \,\mathrm{eV},$ $\Gamma_{\mathrm{O1s}}= 0.13 \,\mathrm{eV},$ $\Gamma_{\mathrm{S1s}}= 0.59 \,\mathrm{eV},$ and $\Gamma_{\mathrm{S2p}}= 0.054 \,\mathrm{eV}.$  We plot also the power spectra for the ultrafast laser pulses used in the simulations here with FWHM in time and frequency of 128 as and 14.2 eV, respectively.  The central frequencies for the N1s, O1s, S1s, and S2p pulses are 406.4 eV, 532.2 eV, 2473.5 eV, and 164.6 eV, respectively.}
 \end{figure}

Simulations were carried out for four core edges: the K-edges of nitrogen, oxygen, and sulfur as well as the L2-edge of sulfur.  Fig. \ref{fig:xanes} shows the calculated x-ray absorption spectra for these four spectral region.  In this plot, we have used a Fermi's Golden Rule expression convoluted with a Lorentzian lineshape with a constant linewidth (in Ref. \onlinecite{zhang:194306}, an energy-dependent linewidth was used to improve agreement with experiment).  Due to the presence of two oxygen atoms in cysteine, the density of core-excited states near the oxygen K-edge is twice as high as those at the  K-edges of nitrogen and sulfur, as can be seen in the stick spectra of Fig. \ref{fig:xanes}. By examining the CI coefficients (not shown here), we see that for any given oxygen core-excitation, the core hole is found to be entirely localized on one oxygen atom or the other. In other words, there is no detectable coupling between the two cores at this level of theory.  The sulfur L2 edge likewise has a higher density of states due to the three 2p orbitals of sulfur.

We assume transform-limited Gaussian pulses, with full-width half max (FWHM) in intensity of 128 as (14.2 eV).  The power spectra for the pulses are shown as blue shaded regions overlaying the XANES spectra in Fig. \ref{fig:xanes}.  The center frequencies were chosen to cover a significant portion of the absorption spectrum, and the large bandwidth can impulsively excite valence-excited states between 5 and 12 eV.  The calculated UV absorption for this model system was given in Ref. \onlinecite{zhang:194306}.

The coupling strength between valence and core excitations depends on the dominant hole-particle orbital pairs of the excitations. If they share the same dominant hole and/or particle orbitals, their coupling is strong and thus the corresponding cross peak is visible in the 2D photon echo or SXRS spectrum; otherwise the cross peak cannot be seen. So by examine the cross peaking pattern in the 2D photo echo or SXRS spectrum, we are able to gain insight about electronic structures of excited states.

\subsection{Four-Wave Mixing}\label{subsec:photonechosim}
\begin{figure*}[htbp]
  \includegraphics[width = 17 cm] {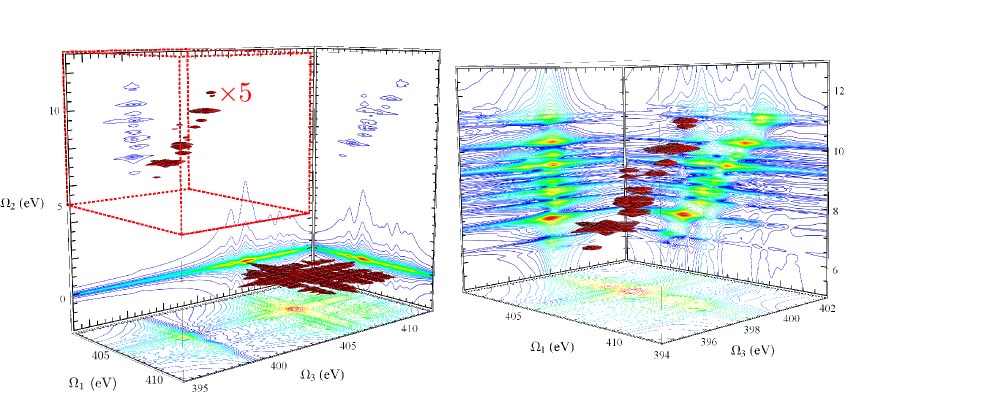}
\caption{\label{fig:kI-N1N1-full} 3D contour plots of $|S_{\ki}|$, the modulus of the $\ki$  photon echo NNNN signal with all four pulses tuned to the nitrogen K-edge, and all pulses polarized parallel (XXXX).  (Left) The full 3D $\kkk_I$ XPE signal.  The portion of the signal enclosed in the red dashed box (multiplied by 5 here to increase visibility) is responsible for the x-ray Raman resonances. (Right) Enlarged spectrum of the region $\Omega_2>0$, highlighted by the red box in the full spectrum. The walls of the 3D box enclosing the contour plot show 2D projections of the 3D data, defined by Eq. \ref{eq:projs}.}
 \end{figure*}
The calculated $\ki$  signal for cysteine with all four pulses tuned to the nitrogen K-edge, the NNNN signal, and polarized parallel to each other (XXXX) is shown as a 3D contour plot in Fig.  \ref{fig:kI-N1N1-full}. $\Omega_1$ and $\Omega_3$ resonances reveal core-excited states (near 400 eV for nitrogen), while $\Omega_2$ covers valence-excitations accessed via the core-excitations.  The most intense peak occurs near $\Omega_2 = 0 \, \textrm{eV}$ (herein labeled the elastic contribution), corresponding to pathways where the second pulse returns the system back to the ground state.  One could, in principle, minimize this contribution by redshifting pulse $\kkk_2$ with respect to pulse $\kkk_1$ such that the second pulse is only resonant with emission from core-excited states to valence-excited states.  However, this is not necessary for this system, since the valence excitations are
spectrally far removed from the elastic contribution (calculated valence excitation frequencies $\omega_{g'g}$ are between 5.7 eV and 11.5 eV).

We focus on the region $ 5 \,\ev \leq \Omega_2 \leq 12 \, \ev$ shown in an expanded scale in the right panel of Fig. \ref{fig:kI-N1N1-full}.  Also shown with the 3D contour plot are 2D projections, displayed as the walls to the bounding box.  These are defined by
\begin{equation}\begin{split}\label{eq:projs}
S_{\ki}(\bar{\Omega}_1, \Omega_2, \Omega_3) &=  \int_{-\infty}^\infty \mathrm{d}\Omega_1 S_{\ki}(\Omega_1, \Omega_2, \Omega_3) \\
S_{\ki}(\Omega_1, \bar{\Omega}_2, \Omega_3) &=  \int_{5 \ev}^{13\ev} \mathrm{d}\Omega_2 S_{\ki}(\Omega_1, \Omega_2, \Omega_3) \\
S_{\ki}(\Omega_1, \Omega_2, \bar{\Omega}_3) &=  \int_{-\infty}^\infty \mathrm{d}\Omega_3 S_{\ki}(\Omega_1, \Omega_2, \Omega_3)
\end{split}\end{equation}
where the region of integration is chosen to exclude the elastic component.
\begin{figure*}[htbp]
  \includegraphics[width = 12.5 cm] {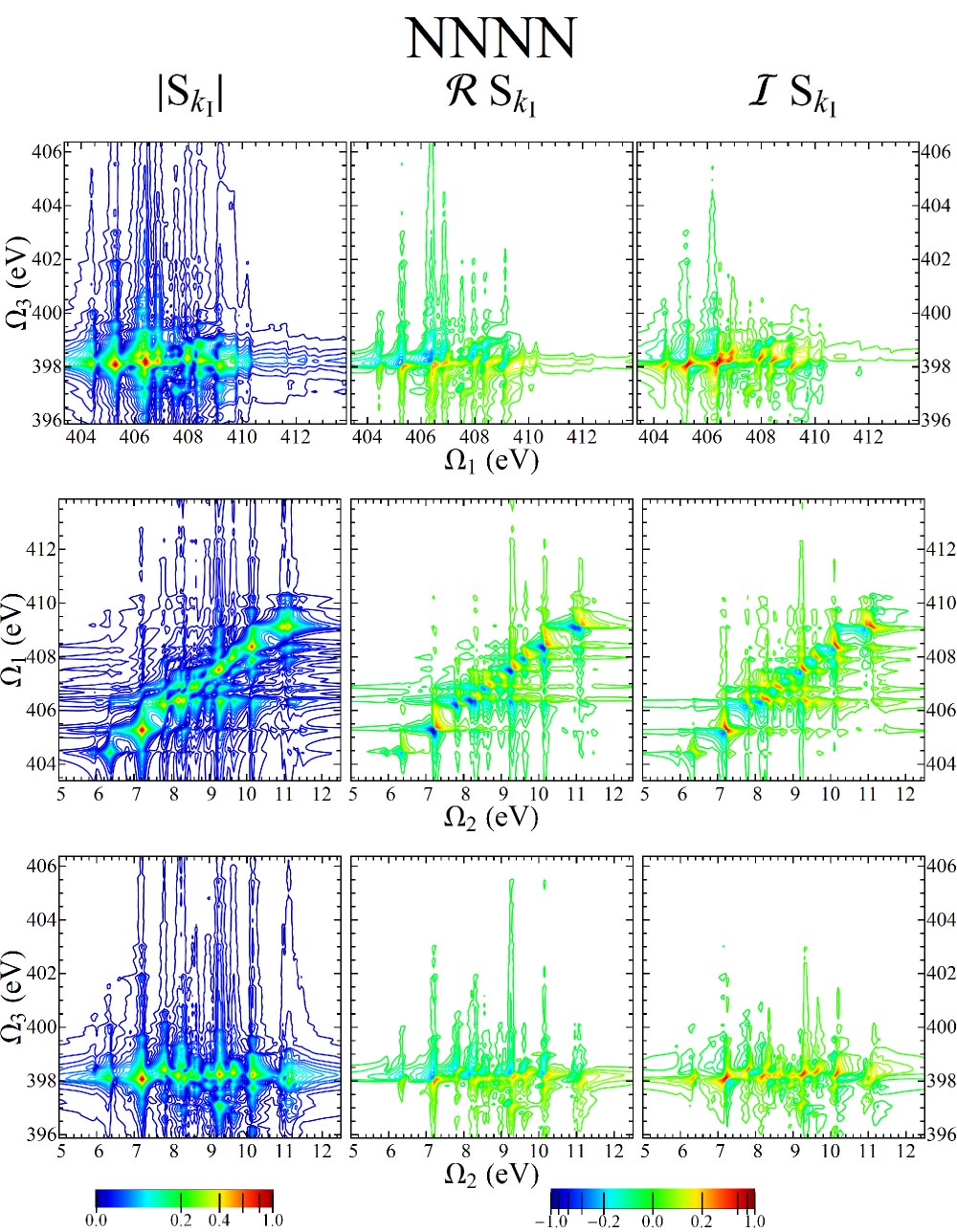}
\caption{\label{fig:kI-NN1}2D projections of the 3D $\ki$ signal of Fig. \ref{fig:kI-N1N1-full}, using an NNNN pulse sequence and XXXX polarization.  Top row: $S_{\ki}(\Omega_1,\bar{\Omega}_2,\Omega_3)$, Middle row: $S_{\ki}(\Omega_1,\Omega_2,\bar{\Omega}_3)$, and Bottom row: $S_{\ki}(\bar{\Omega}_1,\Omega_2,\Omega_3)$.  The left, middle, and right columns show $|S_{\ki}|$, $\Re S_{\ki}$, and $\Im S_{\ki}$, respectively. Each signal is plotted using a nonlinear scale shown in the color bars.}
\end{figure*}

By varying the phase of the fourth pulse it is possible to recover the real and imaginary parts of the time-domain signal prior to the application of the numerical Fourier transform. Fig. \ref{fig:kI-N1N1-full} displays the modulus of Eq. \ref{eq:ki}.  In Fig. \ref{fig:kI-NN1} we show the real, imaginary, and modulus of Eq. \ref{eq:projs}. For these and other 2D contour plots presented here, each signal is individually normalized to lie between -1 and 1, and we use a nonlinear scale\cite{abramavicius_energy-transfer_2010} to highlight weak features in the signal.  The real and imaginary signals contain additional phase information regarding the molecular response.  We look to the $\ki$ signal to explain why certain valence states show up in the SXRS signals, and this information is more easily read from the modulus signal as it does not contain dispersive lineshapes.

The NNNN $\ki$ signal, shown in Figs. \ref{fig:kI-N1N1-full} and \ref{fig:kI-NN1}, shows that the presence of a nitrogen core hole does not greatly effect the valence orbitals. For every core excitation we see in $\Omega_1$, there is a single valence excitation to which it is strongly coupled.  Further, these peaks fall mainly along a diagonal line where $\Omega_1$ is equal to $\Omega_2 + \sim 398$ eV.  This corresponds roughly to the energy difference between the highest occupied molecular orbital (HOMO) and the molecular orbital (MO) which corresponds to the nitrogen 1s core orbital.  This is seen most clearly in the $\Omega_3/\Omega_2$ plot in the bottom row of Fig. \ref{fig:kI-NN1}.  Peaks in $\Omega_3$ correspond to the energy difference between core and valence levels, and we see here that this difference is 398 eV independent of $\Omega_2$.  We find roughly a one-to-one correspondence between valence excitations and nitrogen core excitations.  We use a picture in which the core electron is first promoted to a single unoccupied MO (or a CI expansion thereof), following which an electron from the HOMO falls down to fill the vacant core hole.  A nitrogen core hole does not appreciably change the excited particle states from those available to valence electrons, and the diagonal and horizontal features of the middle and lower rows of Fig. \ref{fig:kI-NN1} show this clearly.

\begin{figure*}[htbp]
  \includegraphics[width = 12.5 cm] {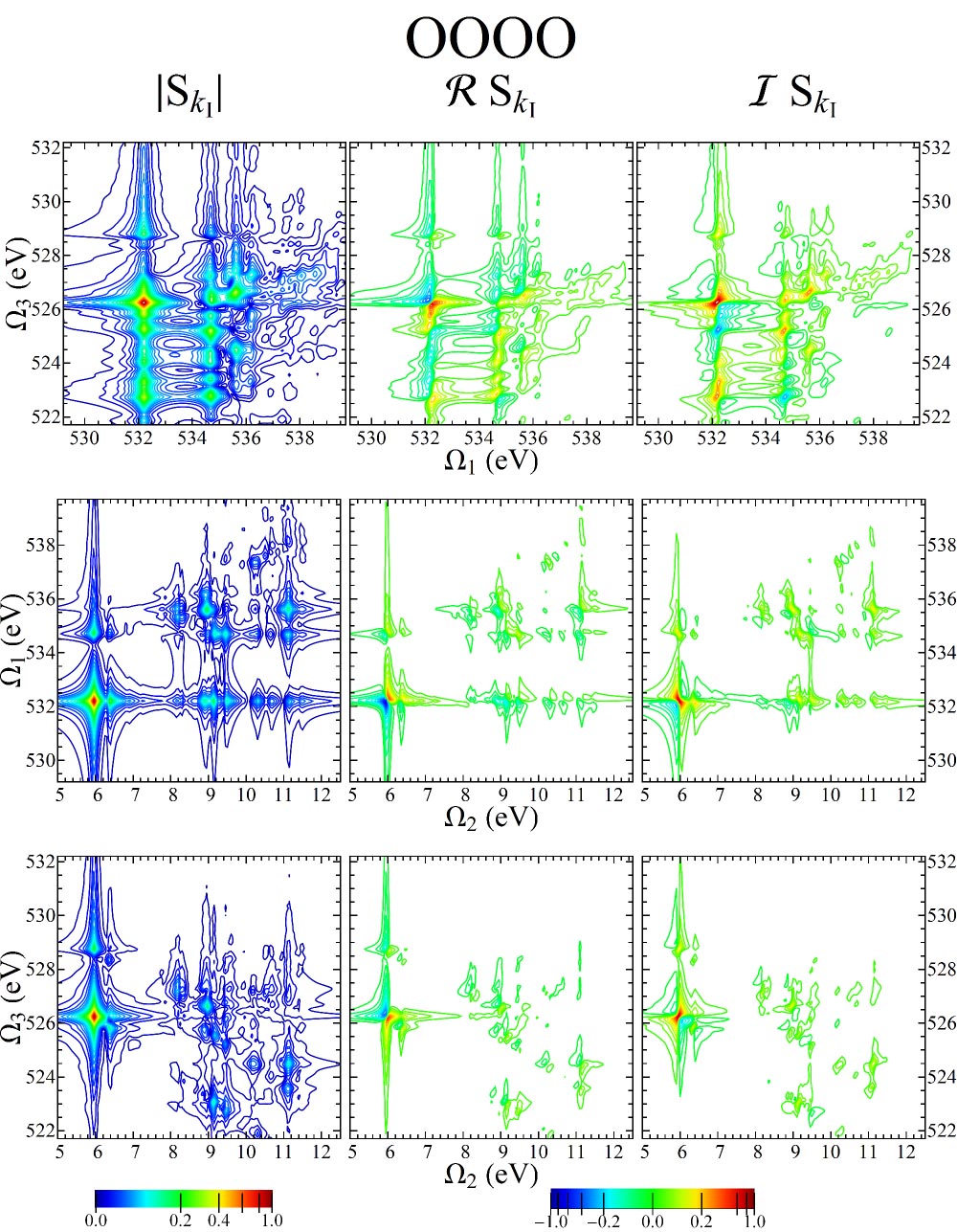}
\caption{\label{fig:kI-OO1}Same as Fig. \ref{fig:kI-NN1}, using an OOOO pulse sequence and XXXX polarization.}
\end{figure*}

In Fig. \ref{fig:kI-OO1} we show the OOOO $\ki$ signal.  In this case, there is no one-to-one correspondence between valence and core excitations.  Rather, a given core excitation is projected onto many different valence excitations.  This indicates that a core hole on one of the oxygen atoms greatly effects the valence MOs, causing considerable orbital rotation which is not the case for a nitrogen core hole.

The NNNN and OOOO photon echo signals described above are single-color experiments.  As we can see from Fig. \ref{fig:kI-NN1}, the information regarding coupling between the core- and valence-excited manifold in the $\Omega_1$ and $\Omega_3$ dimensions is largely the same.  A two-color configuration, where two different core orbitals are accessed, allows to probe valence-core coupling at different locations.  There are three possible combinations of four pulses of two different colors A and B: AABB, ABBA, and ABAB.  In Ref. \onlinecite{schweigert_probing_2008}, Schweigert and Mukamel looked at the $\ki$ signal from different isomers of aminophenol using the AABB and ABAB configuration.  Here we focus on the AABB configuration, as it is the only one that results in a valence-excited coherence during the second time interval.  Ref. \onlinecite{schweigert_probing_2008} presents $\ki$ signals for constant $t_2=0$ rather than Fourier transforming with respect to $t_2$.
\begin{figure*}[htbp]
  \includegraphics[width = 12.5 cm] {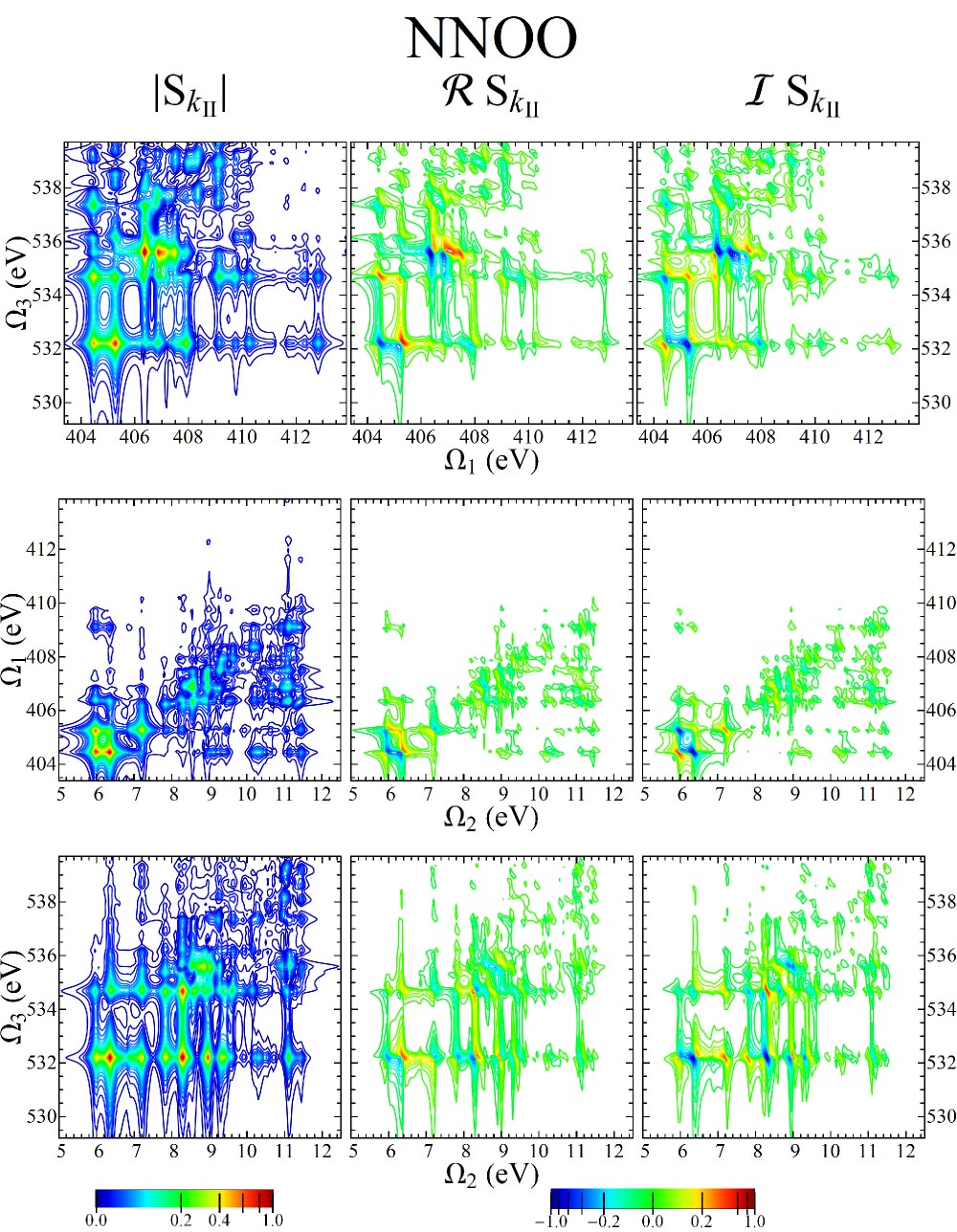}
\caption{\label{fig:kII-NO} 2D projections of the 3D $\kii$ signal using an NNOO pulse sequence with XXXX polarization.}
 \end{figure*}

Fig. \ref{fig:kII-NO} shows the 2D projections of the NNOO $\kii$ signal.  The major difference between the $\ki$ and $\kii$ signals, apart from their response to inhomogeneous broadening which we do not treat here, is that $\Omega_3$ resonances occur at the core-excitation energies $\omega_{eg}$ rather than at$\omega_{eg'}$.  This is due to the fact that the first and second pulse pair act on the same side of the density matrix, which makes the $\kii$ signal more straightforward.  The bottom row of Fig. \ref{fig:kII-NO} shows $|S_{\kii}(\bar{\Omega}_1,\Omega_2,\Omega_3)|$, and makes it clear that the coupling between the valence-excited and oxygen core-excited manifolds is much different than that for a nitrogen core excitation.  The first few oxygen core-excited states, with energies between 532 eV and 535 eV, are each coupled to many different valence excitations.  The $|S_{\ki}(\Omega_1,\Omega_2,\bar{\Omega_3})|$ plot in the middle row of Fig. \ref{fig:kII-NO} is similar to the same
plot from Fig. \ref{fig:kI-NN1}.  However there are distinct differences.  The valence-excited state at 5.74 eV is absent from the NNNN signal, but is strong in the NNOO signal.  Furthermore, in the NNOO $\kii$ signal, we see that this state is coupled to multiple nitrogen core-excited states, in contrast to NNNN signal where there was a one-to-one correspondence between the two manifolds.
\begin{figure*}[htbp]
  \includegraphics[width = 12.5 cm] {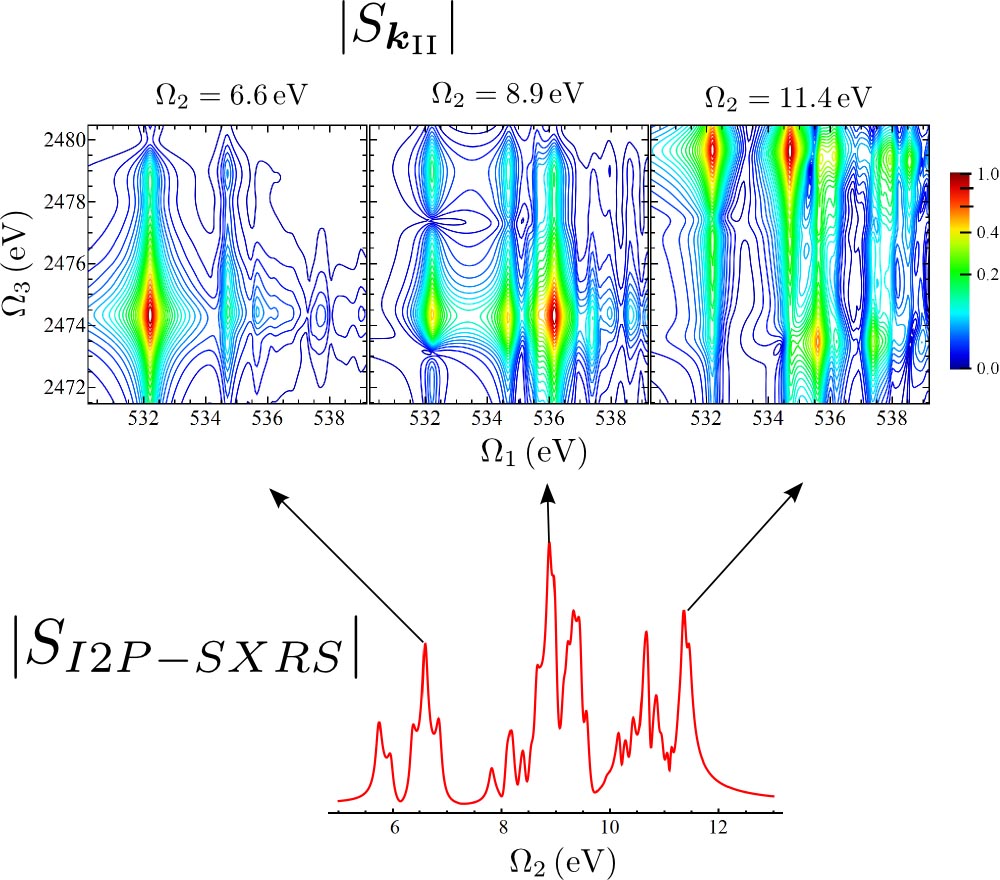}
\caption{\label{fig:kIi-slices} (Top) Constant-$\Omega_2$ slices of the 3D $\kii$ signal using an OOSS pulse sequence, XXXX polarization.  Specifically, we plot $|S_{\kii}(\Omega_1,6.6 \, \ev,\Omega_3)|$, $|S_{\kii}(\Omega_1,8.9 \, \ev,\Omega_3)|$, and $|S_{\kii}(\Omega_1,11.4 \, \ev,\Omega_3)|$. (Bottom) The I2P-SXRS signal using an OS pulse sequence with XX polarization.}
 \end{figure*}

The projections defined in Eq. \ref{eq:projs}, and displayed in Figs. \ref{fig:kI-NN1} and \ref{fig:kII-NO} provide only one example of how the full 3D spectrum can be parsed into 2D spectra, which are more easily interpreted.  Another method, one which will give direct insight into the origin of the two-color I2P-SXRS resonances, is to display 2D $\Omega_1/\Omega_3$ plots for constant $\Omega_2$.  In Fig. \ref{fig:kIi-slices} we show slices of the OOSS $\kii$ signal for constant $\Omega_2$ corresponding to three different peaks in the OS I2P-SXRS spectrum.  These give a correlation plot between core-excited states located at different atomic centers, corresponding to a specific valence-excited state.  We see that the valence-excitations at 6.6 eV and 8.9 eV have are coupled to the same sulfur 1s core-excited states, but different oxygen core-excited states.  The higher-energy 11.4 eV valence excitation, on the other hand, has a much different excitation pattern, being mainly coupled to higher-lying sulfur
excited-states.

Finally we note that the four-wave mixing signals can be used to disentangle the effects of pulse polarization in x-ray Raman signals.  Polarization is more important in x-ray Raman than in traditional optical Raman.  In a vibrational resonant Raman experiment, the system is transiently promoted to an electronically-excited state before de-excitation to a vibrationally excited state in the ground electronic state.  Under the Condon approximation, the transition dipoles for the upward and downward transitions will always be parallel to each other.  In x-ray Raman spectroscopy, there is no constraint on the angle between the upward $\boldsymbol{V}_{eg}$ and downward $\boldsymbol{V}_{eg'}$ transition dipoles.  In general, for a third-order nonlinear experiment such as the $\ki$ and $\kii$, there is an orientational factor $\mathcal{I}^{\boldsymbol{e}_1,\boldsymbol{e}_2,\boldsymbol{e}_3, \boldsymbol{e}_4}_{\boldsymbol{V}_{e'g'},\boldsymbol{V}_{ge'},\boldsymbol{V}_{ge}
\boldsymbol{V}_{eg'}}$ multiplying each term in Eqs. \ref{eq:ki} and \ref{eq:kii}, defined in Ref. \onlinecite{sma_reference}.  By setting the angles
\begin{equation}\label{eq:pol1}
(\boldsymbol{e}_1,\boldsymbol{e}_2,\boldsymbol{e}_3,\boldsymbol{e}_4) = (0^{\circ},0^{\circ},\theta_\mathrm{MA},\theta_\mathrm{MA})
\end{equation}
where V is vertical and $\theta_\mathrm{MA}=54.7^{\circ}$ is the magic angle, the orientational factor becomes equal to $(\boldsymbol{V}_{e'g'}\cdot\boldsymbol{V}_{ge'})(\boldsymbol{V}_{eg'}\cdot\boldsymbol{V}_{ge})$. This is one of the three intrinsic isotropic signal components. This polarization configuration corresponds to the 2P-SXRS signal when the pump and probe are polarized at the magic angle to each other.
\begin{figure*}[htbp]
  \includegraphics[width = 12.5 cm] {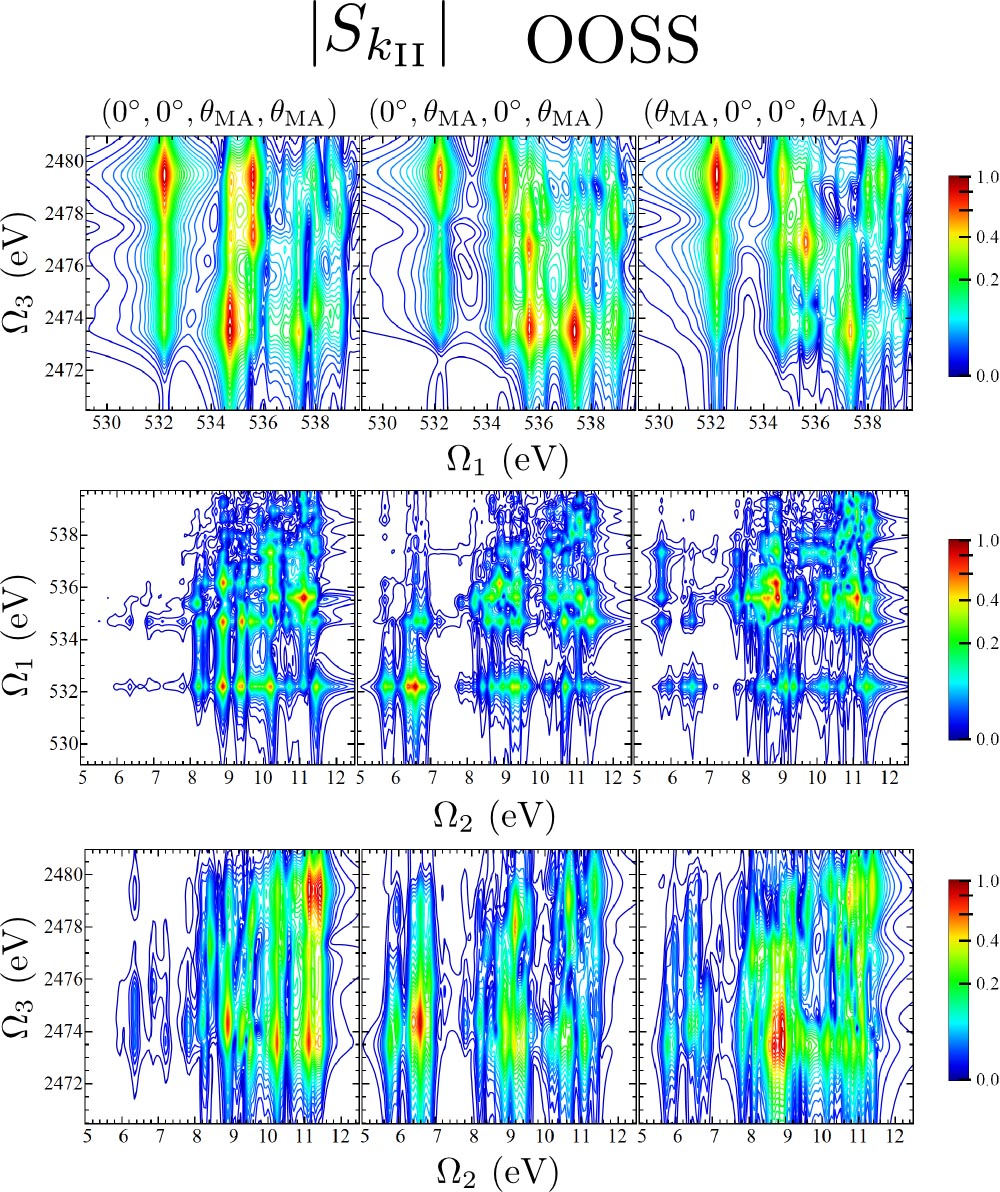}
\caption{\label{fig:kIi-polarization} $\kii$ signals simulated for an OOSS pulse sequence for three different polarization configurations.  Each column represents a different polarization schemes, with the polarization vectors of pulses 1 through 4 indicated at the top of each column.  The top row shows $|S_{\kii}(\Omega_1,\bar{\Omega}_2,\Omega_3)|$, the middle row $|S_{\kii}(\Omega_1,\Omega_2,\bar{\Omega}_3)|$, and the bottom row $|S_{\kii}(\bar{\Omega}_1,\Omega_2,\Omega_3)|.$}
 \end{figure*}
It is possible to isolate all three of the isotropic components.  When
\begin{equation}\label{eq:pol2}
(\boldsymbol{e}_1,\boldsymbol{e}_2,\boldsymbol{e}_3,\boldsymbol{e}_4) = (0^{\circ},\theta_\mathrm{MA},0^{\circ},\theta_\mathrm{MA}),
\end{equation}
the orientational factor becomes $(\boldsymbol{V}_{e'g'}\cdot\boldsymbol{V}_{eg'})(\boldsymbol{V}_{ge'}\cdot\boldsymbol{V}_{ge})$.  And finally, when
\begin{equation}\label{eq:pol3}
(\boldsymbol{e}_1,\boldsymbol{e}_2,\boldsymbol{e}_3,\boldsymbol{e}_4) = (0^{\circ},\theta_\mathrm{MA},\theta_\mathrm{MA},0^{\circ})
\end{equation}
the orientational factor becomes $(\boldsymbol{V}_{e'g'}\cdot\boldsymbol{V}_{ge})(\boldsymbol{V}_{eg'}\cdot\boldsymbol{V}_{ge'})$.
The signal resulting from an arbitrary pulse polarization configuration can be written as a linear combination of these three isotropically averaged components.  Fig. \ref{fig:kIi-polarization} depicts the $\kii$ signal, using an OOSS pulse sequence, for the three field polarization configurations described here.

These polarized signals show distinct differences related to the angles between the various transition dipoles involved.  For example, the largest peak in the $(0^{\circ},\theta_\mathrm{MA},0^{\circ},\theta_\mathrm{MA})$ signal, shown in the middle column of Fig. \ref{fig:kIi-polarization}, occurs at $(\Omega_1,\Omega_2,\Omega_3)=(532 \, \textrm{eV},6.6 \, \textrm{eV}, 2474 \, \textrm{eV})$ and is mostly absent from the $(0^{\circ},0^{\circ},\theta_\mathrm{MA},\theta_\mathrm{MA})$ signal, shown in the left column.  This indicates that the transition dipoles connecting the ground state to the given core-excited states are nearly perpendicular to the transition dipoles between these same core states and the valence-excited state at 6.6 eV.  The polarization dependence of broadband x-ray signals can be used as an experimental check on the inter-dipole angles predicted by electronic structure calculations.

\FloatBarrier
\subsection{Dispersed versus Integrated Two-Pulse SXRS}\label{subsubsec:fdsxrsSIM}

\begin{figure}[htbp]
  \centering
    \includegraphics[width = 8.5 cm] {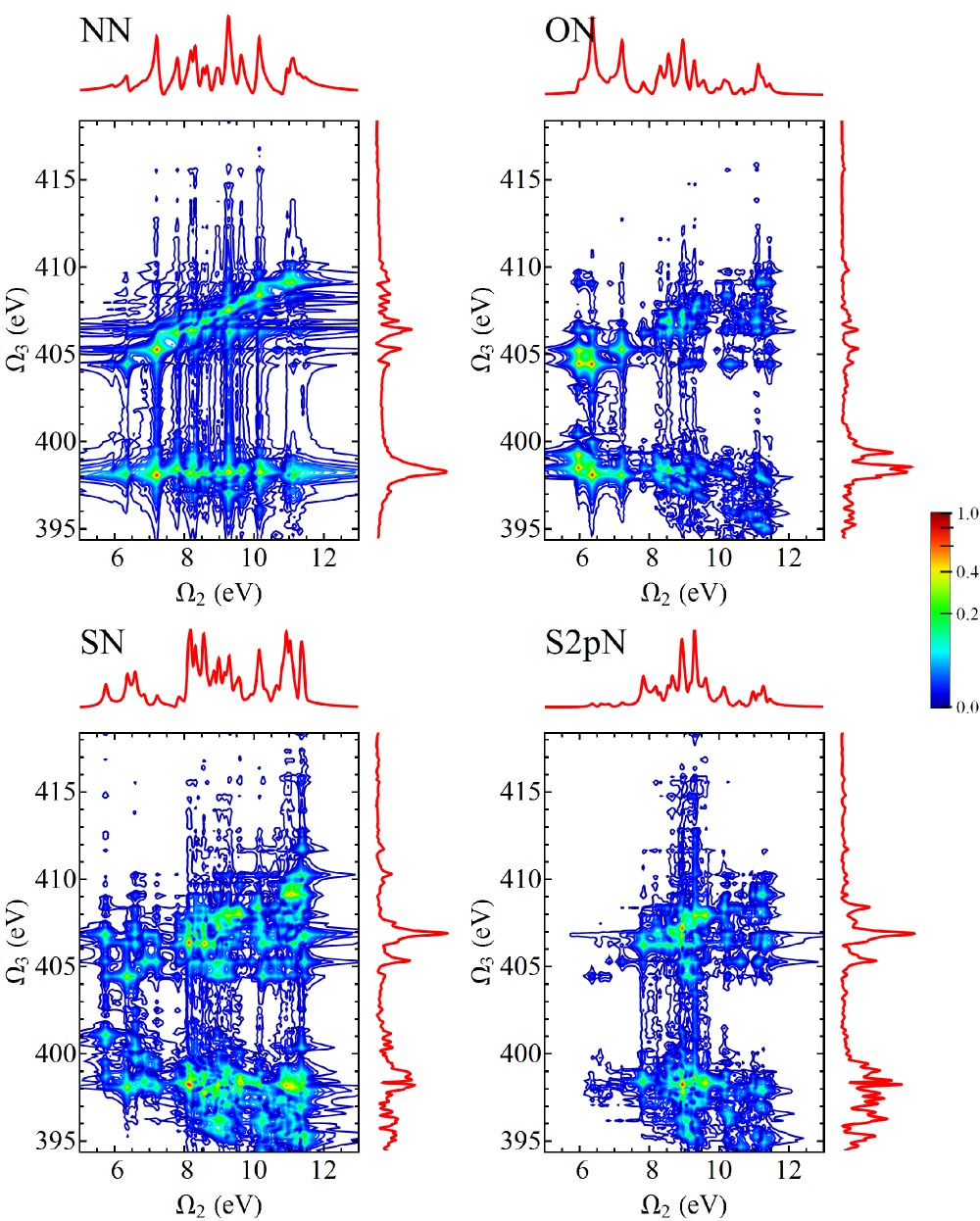}
  \caption{\label{fig:sxrsN1sProbe} Frequency-dispersed SXRS spectra (Eq. \ref{eq:fd1})for cysteine, for various pump pulse detunings with the probe tuned to the N1s-edge transition frequency.  The vertical axis is the dispersed frequency $\Omega_3$, and the horizontal axis is the Fourier conjugate of the interpulse delay $\Omega_2$. The 2D signal is plotted using an arcsinh nonlinear scale, and the projections are shown, in a linear scale, in the top and right marginals.  The trace along the top corresponds to the integrated SXRS signal.}
\end{figure}
The frequency-dispersed two-pulse SXRS signal reveals information on the coupling between valence excitations, excited impulsively by the pump pulse, and the core-excitations resonant with the probe pulse.  The integrated SXRS signals for cysteine were presented in Ref. \onlinecite{zhang:194306}.

\begin{figure}[htbp]
  \centering
    \includegraphics[width = 8.5 cm] {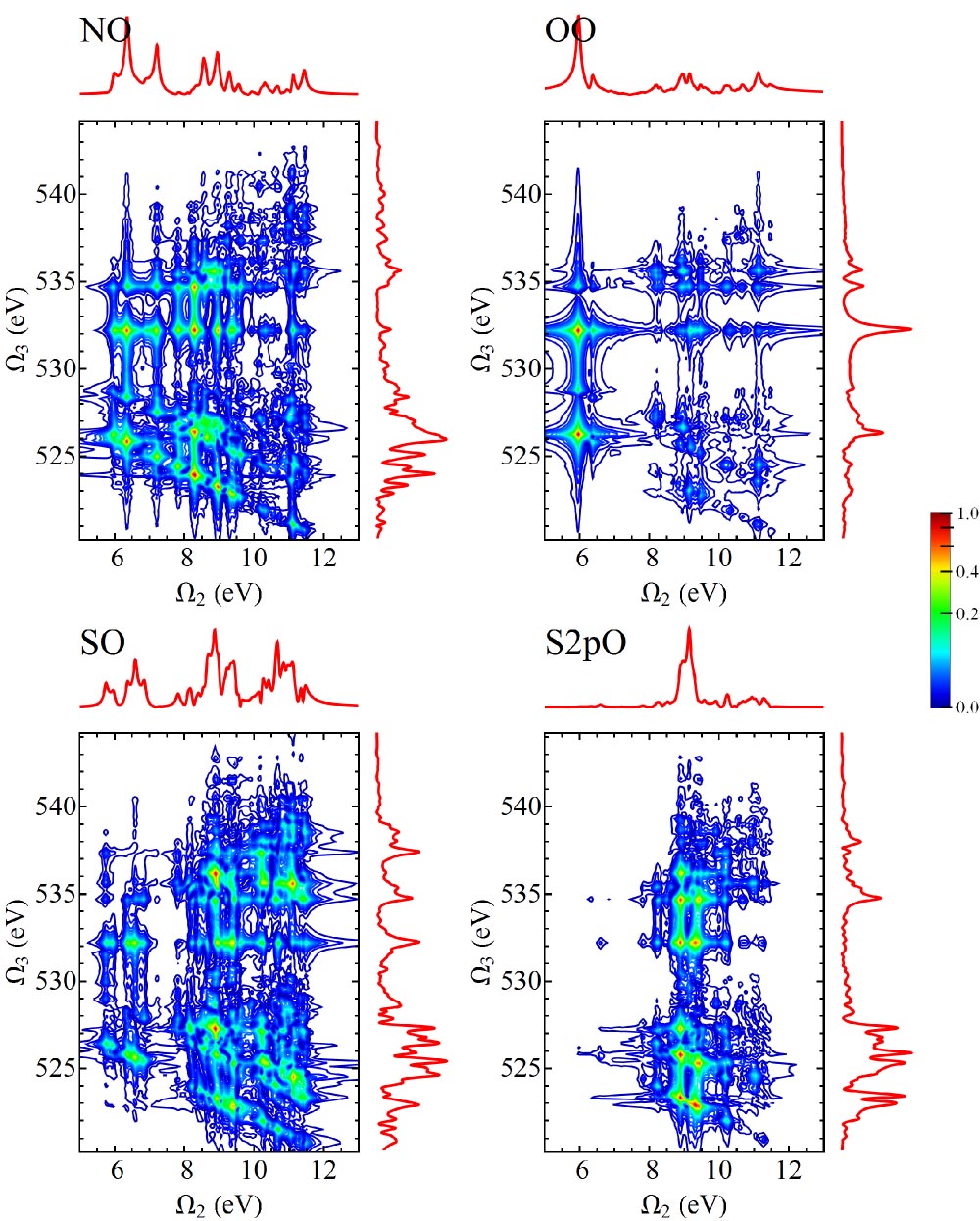}
  \caption{\label{fig:sxrsO1sProbe} Same as Fig. \ref{fig:sxrsN1sProbe}, but with the probe tuned to the oxygen K-edge.}
\end{figure}

\begin{figure}[htbp]
  \centering
    \includegraphics[width = 8.5 cm] {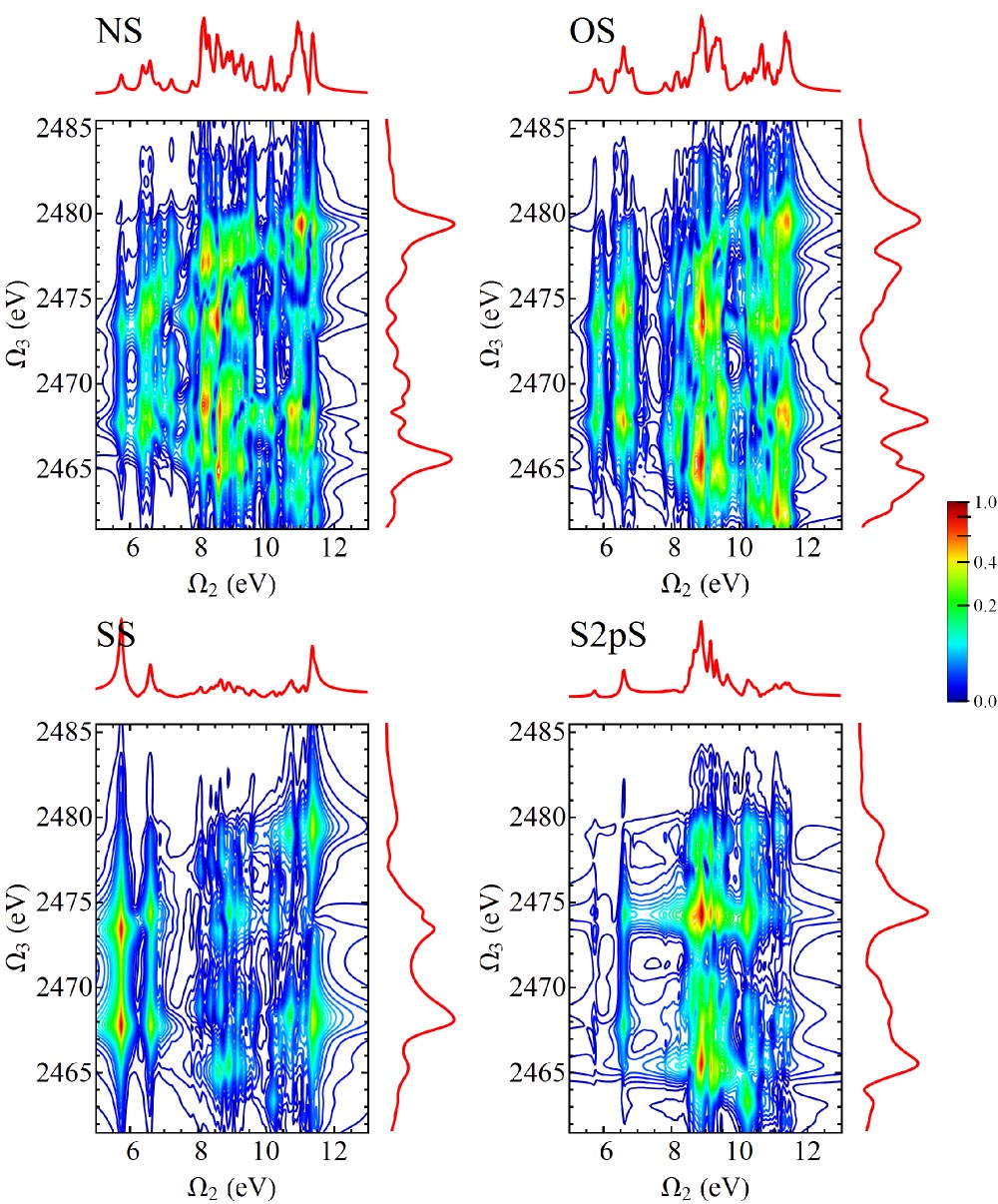}
  \caption{\label{fig:sxrsS1sProbe} Same as Fig. \ref{fig:sxrsN1sProbe}, but with the probe tuned to the sulfur K-edge.}
\end{figure}

\begin{figure}[htbp]
  \centering
    \includegraphics[width = 8.5 cm] {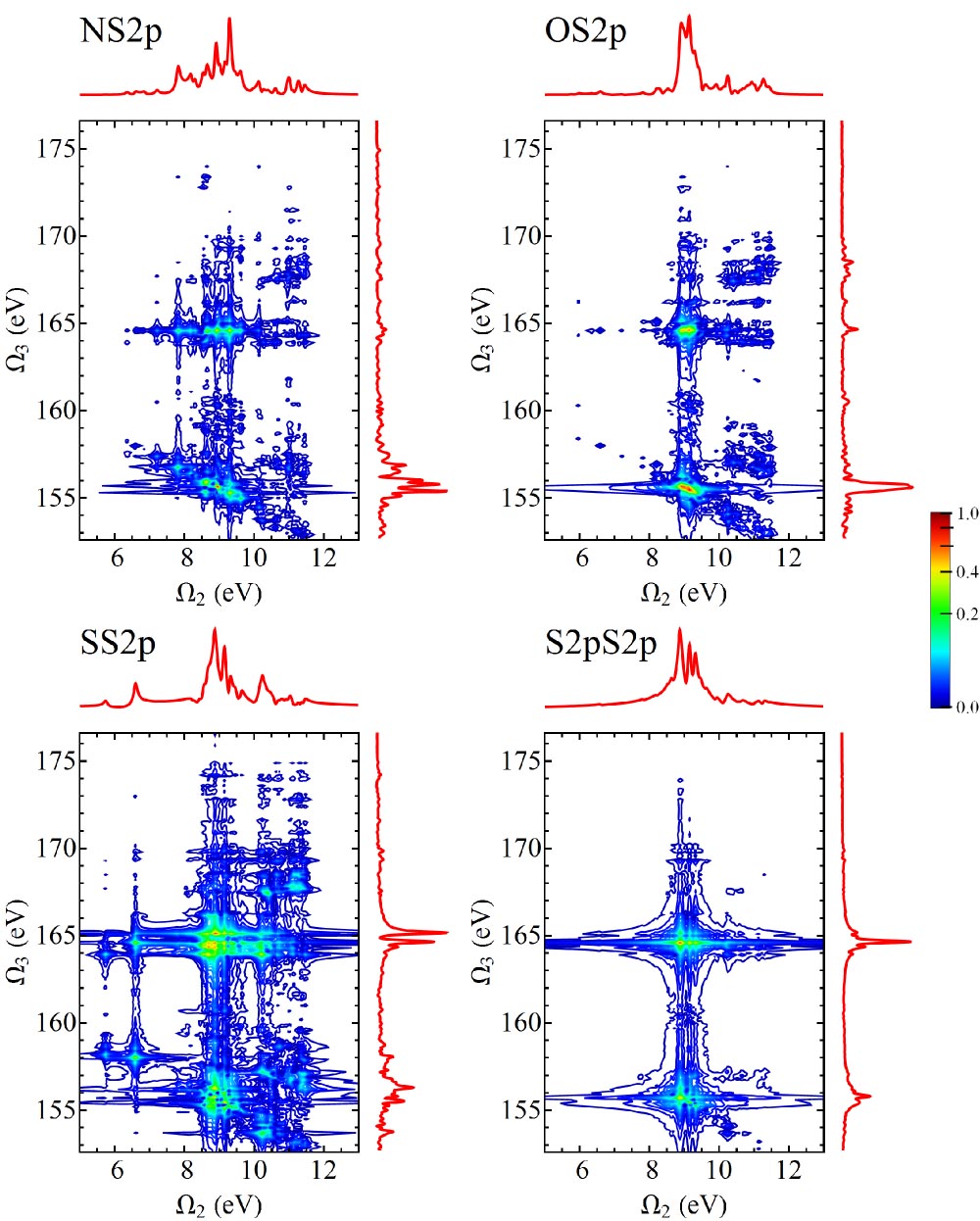}
    \caption{\label{fig:sxrsS2pProbe} Same as Fig. \ref{fig:sxrsN1sProbe}, but with the probe tuned to the sulfur L-edge.}
\end{figure}

In Fig. \ref{fig:sxrsN1sProbe} we show the D2P-SXRS signals, using all-parallel polarization, using a probe pulse tuned to the nitrogen K-edge and pump pulses tuned to the four core edges considered here.  We can see from Eq. \ref{eq:fd1} that any given resonance between a valence-excited state $g'$ and a core-excited state $e$, will generate two peaks located at $(\Omega_2,\Omega_3) = (\omega_{g'g},\omega_{eg})$ and $(\omega_{g'g},\omega_{eg'})$.  Because our model does not possess valence excitations with energies below 5.74 eV, it is easy to distinguish between these types of peaks.  In Figs. \ref{fig:sxrsO1sProbe} through \ref{fig:sxrsS2pProbe} we show the remaining D2P-SXRS signals for all other probe detunings.  Along with the 2D contour plots, we also show the 1D projections along each axis, defined analogously to Eq. \ref{eq:projs}. The topmost projection is the integrated 2P-SXRS signal.


\begin{figure}[htbp]
  \includegraphics[width = 8.5 cm] {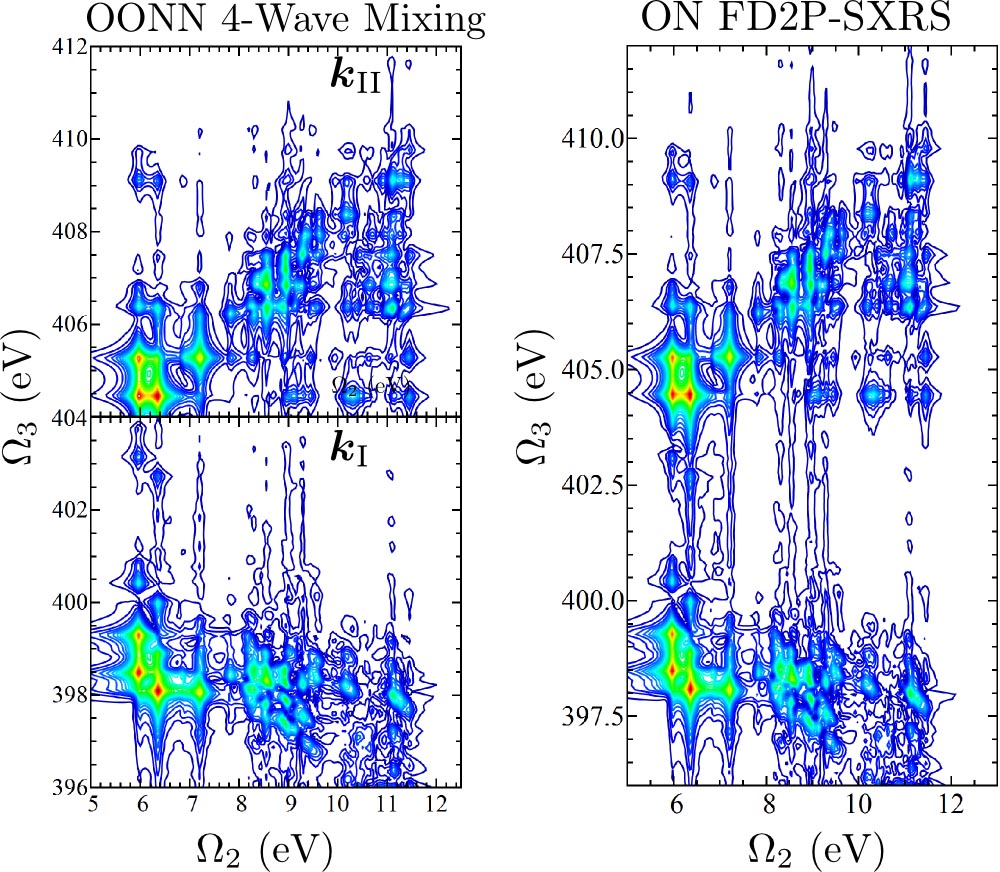}
\caption{\label{fig:photecho_comp} (Left column) Four-wave mixing signals using the OONN pulse sequence (XXXX polarized).  On top is $|S_{\kii}(\bar{\Omega_1},\Omega_2,\Omega_3)|$ and on the bottom is $|S_{\ki}(\bar{\Omega_1},\Omega_2,\Omega_3)|$. (Right column) The modulus D2P-SXRS signal with an ON pulse sequence, XX polarized.}
\end{figure}

In Fig. \ref{fig:photecho_comp} we compare the $\ki$ and $\kii$ signals with the dispersed Raman signals, both taken with a two-color setup.  If the $\Omega_1$ axis is integrated out, formally equivalent to setting $t_1=0$, then the pump-probe signal gives identical information to the experimentally more challenging $\ki$ and $\kii$ signals.  However, in systems which feature both low- and high-energy valence excitations, where the $\omega_{eg}$ and $\omega_{eg'}$ peaks would overlap, the $\ki$ and $\kii$ signals would be able to more clearly identify resonances between core- and valence-excited states.  Furthermore, four-wave mixing gives direct cross-correlations between core-excitations on different atoms while this information must be inferred from the pump-probe measurements.
\FloatBarrier
\subsection{Dispersed versus Integrated Three-Pulse SXRS}\label{subsubsec:fdsxrsSIM3pulse}
\begin{figure}[htbp]
  \centering
    \includegraphics[width = 8.5 cm] {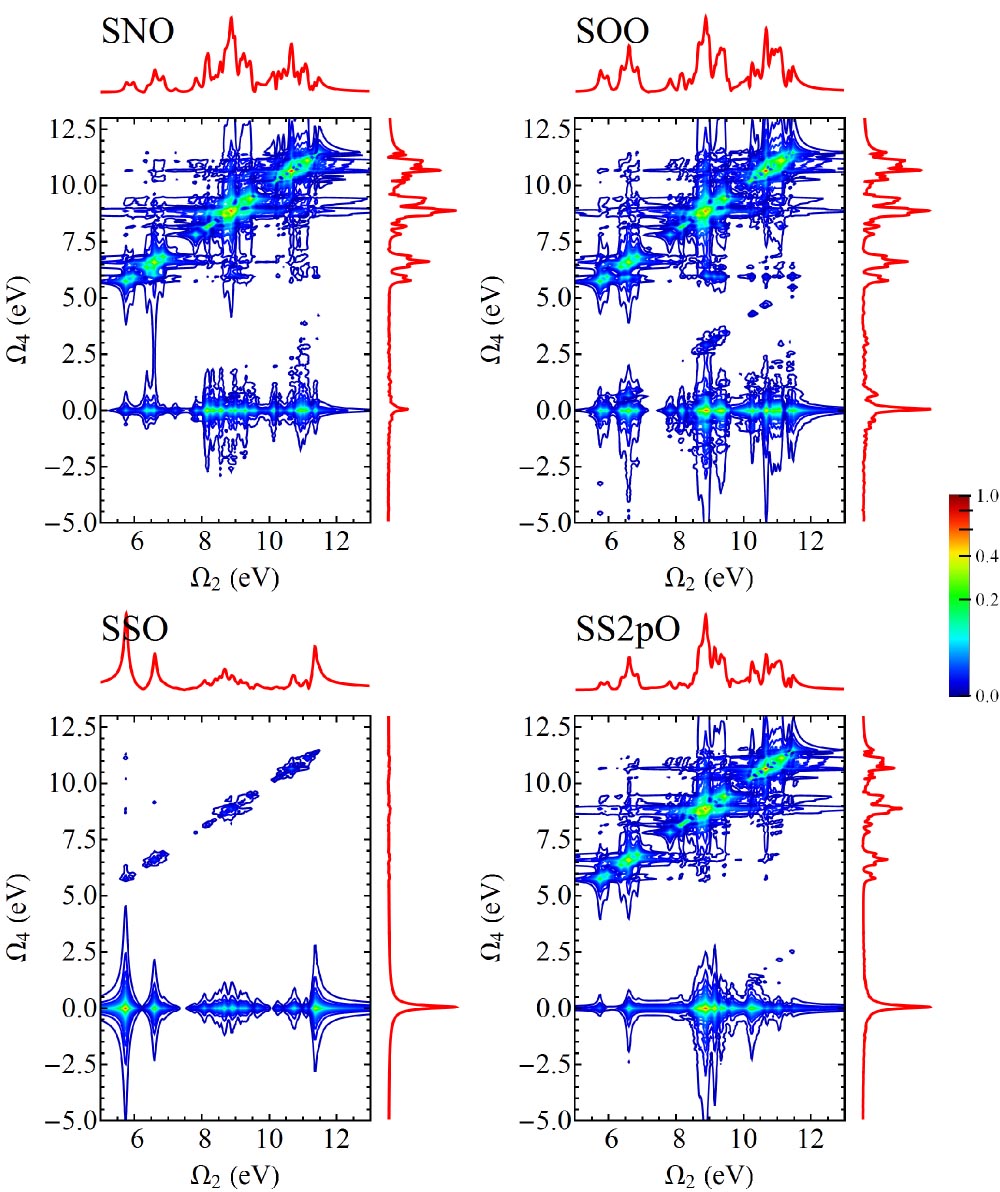}
  \caption{\label{fig:2DsxrsO1sProbe} I3P-SXRS signals from cysteine with the first  pulse  resonant with the S1s transition and the third at the O1s transition, with XXX polarization.}
\end{figure}

\begin{figure}[htbp]
  \includegraphics[width = 8.5 cm] {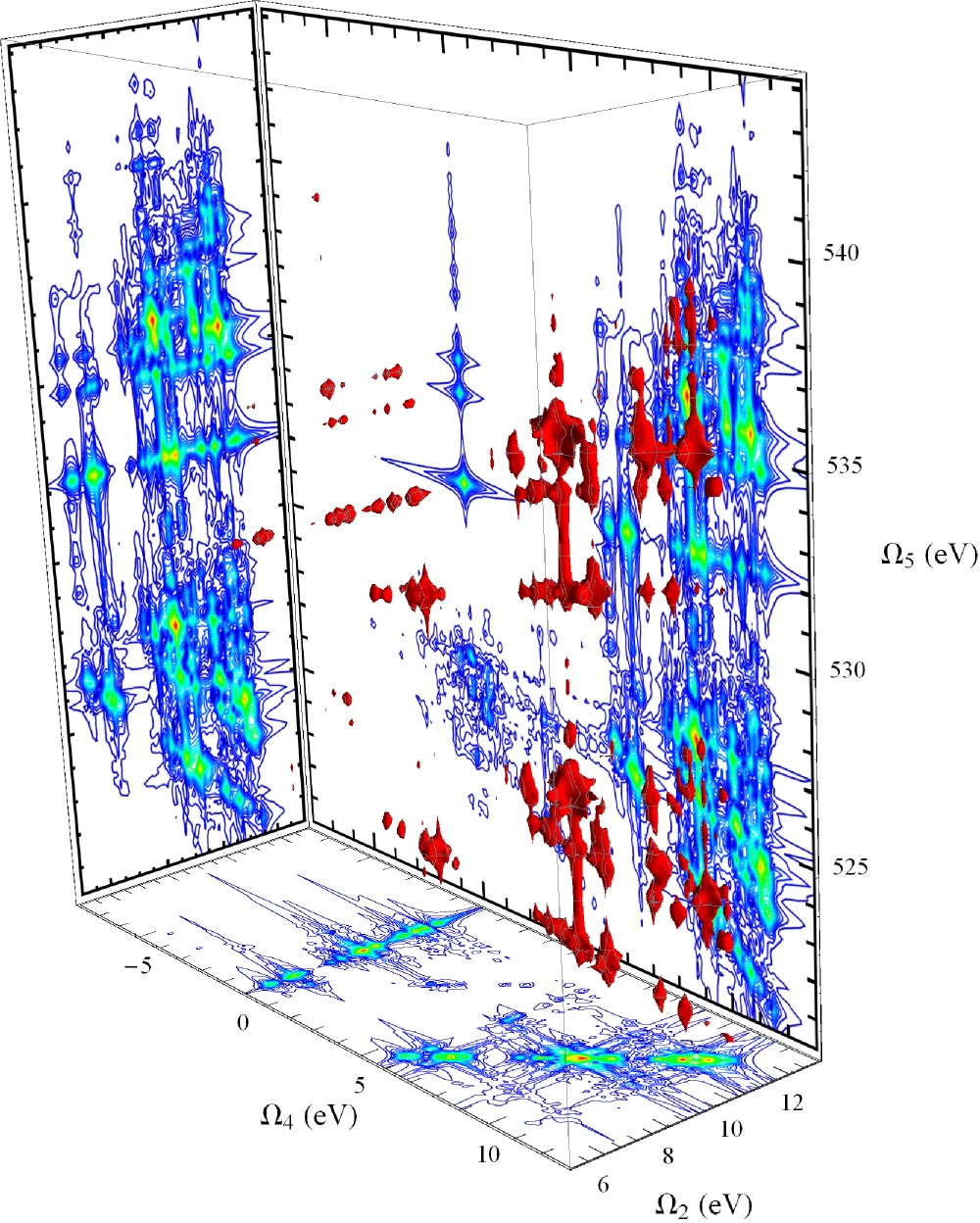}
\caption{\label{fig:FD3Psxrs} D3P-SXRS signals from cysteine using an SOO pulse configuration, with XXX polarization.  $\Omega_2$ and $\Omega_4$ are valence excitations, and $\Omega_5$ represents core excitation.}
 \end{figure}

In Fig. \ref{fig:2DsxrsO1sProbe} we show four examples of the I3P-SXRS signals for cysteine, with the first pulse resonant with the sulfur K-edge and the third pulse at the oxygen K-edge.  The signals differ in the tuning of the second pulse.  Diagonal contributions to the signal, where $\Omega_2=\Omega_4$, are largely insensitive to the identity of the middle pulse, while the trace along $\Omega_4 = 0$ are insensitive to the probe pulse.  That is, for an integrated 3-pulse SXRS signal with an ABC pulse configuration (where A,B, and C represent different cores to be excited), the diagonal trace will be similar to the I2P-SXRS signal with an AC pulse configuration and the $\Omega_4 = 0$ trace equal to the I2P-SXRS signal with an AB pulse sequence.  In Fig. \ref{fig:FD3Psxrs} we show the SOO D3P-SXRS signal.

As can be seen in Fig. \ref{fig:2}, the three-pulse Raman signals involve core-excited states $e,$ $e',$ and $e''$.  $e$ and $e'$ contribute to $\alpha_{g'g}$ and $\alpha_{gg''}$ (Eq. \ref{eq:2dsxrs1}), respectively, whereas $e''$ contributes to $\alpha_{g''g'}$ (Eq. \ref{eq:2dsxrs1}).  In our current approximation, core excited states are given by $\ket{e} = c^\dagger_a c_n \ket{g}$ where $c^\dagger$ and $c$ are the Fermi creation and annihilation operators, and $a$ and $n$ are virtual and core orbitals, respectively.  While this is a good representation for $e$ and $e'$, additional contributions $\ket{e''} = c^\dagger_b c^\dagger_a c_i c_n \ket{g},$ where $i$ and $b$ are valence and virtual orbitals, respectively, should be taken into account for $e''$.  These double excitations are neglected in the present simulation, and should result in additional peaks.

\section{Conclusions}
We have shown how the $\ki$ and $\kii$ x-ray four-wave mixing techniques can reveal the coupling between core- and valence-excited states.  Furthermore, it is possible to isolate the three independent rotationally averaged contributions to the signal in these experiments.

The frequency-dispersed two-pulse SXRS signal, a much simpler experiment, gives the same information as the photon-echo provided the first time delay is set to zero.  The D2P-SXRS allows to look inside the probe polarizability, while the four-wave mixing reveals both the pump and probe polarizability.

The time-domain Raman signals have have a further advantage over their frequency-domain counterpart, RIXS, due to their directionality.  In a RIXS experiment the spontaneous signal is not phase-matched and will be emitted evenly over  a $4\pi$ solid angle, only a fraction of which will be collected.  In an SXRS experiment the signal is emitted exclusively in the probe direction, so the entire signal is recovered.  This advantage applies to the $\ki$ and $\kii$ signals as well, where the signal is emitted in a background-free direction.

We have also proposed extending the three-pulse SXRS experiment by adding frequency-dispersed detection.  An example of this three-dimensional fifth-order signal was given.  However, the REW-TDDFT method used to calculate core-excited states, while sufficient for the third-order echo and pump-probe, cannot represent states with both core- and valance-holes.  The development of quantum chemical methodologies more suited to this experiment is called for.

The XFEL light sources are very bright (e.g. $10^{12}$ photons in 85 attoseconds at frequency 8~keV);\cite{emmap._first_2010} the first experiments
performed at the LCLS saturated the core x-ray excitation in a
molecular beam of nitrogen gas.\cite{hoener_ultraintense_2010}
This shows that the stimulated Raman process at these intensities can compete with the Auger process which may be detected.
The strength of the nonlinear signal has been estimated in Ref. \onlinecite{patterson_resource_2010}
by treating the field semiclassically.

\begin{acknowledgments}
  The support of the Chemical Sciences, Geosciences and Biosciences
  Division, Office of Basic Energy Sciences, Office of Science,
  U.S. Department of Energy is gratefully acknowledged. We also
  gratefully acknowledge the support of the National Science
  Foundation (Grant CHE-1058791), and the National Institutes of
  Health (Grant GM-59230). Help on the REW-TDDFT calculations from Niranjan Govind of
  the Pacific Northwest National Lab (PNNL) is greatly appreciated.
\end{acknowledgments}

\appendix
\begin{widetext}
\section{Four-pulse Phase-Matched $\ki$ and $\kii$ Signals}\label{app:4wm}

In this four-photon
stimulated process (Fig. \ref{fig:XPE-diags}), the molecule interacts with four short pulses, described by
\begin{equation}
  \boldsymbol{E}(\mathbf{r},t) = \sum_j \mathbf{e}_j \mathcal{E}_j(t-\tau_j)
  e^{i \mathbf{k}_j \cdot \mathbf{r}} + \textrm{c.c.} \,
\end{equation}
where $\mathbf{e}_i,\mathcal{E}_j,\kkk_j$ and $\tau_j$ are the pulse polarization vector, temporal envelope, wavevector, and arrival time, respectively.
The signal is given by
the change in transmission of the $\kkk_4$ pulse induced by the other
pulses, and is recorded versus the three interpulse delays, defined by $t_1 = \tau_2-\tau_1$, etc. (see Fig.\ref{fig:XPE-diags}).

By placing the detector in a particular wave-vector matching direction, it is possible to selectively probe a subset of the pathways which contribute to the overall molecular response function.
In this paper, we look at the $\ki$ and $\kii$ signals, where we set $\kkk_4$ equal to $\ki=-\kkk_1+\kkk_2+\kkk_3$ and $\kii=\kkk_1-\kkk_2+\kkk_3$, respectively.  The $\kii$ signal can be obtained by using the same experimental geometry as in the $\ki$ signal, if the order of pulses 1 and 2 are interchanged.

We calculate the $\ki$ and $\kii$ signals using time-dependent perturbation theory
with the interaction Hamiltonian
\begin{equation}
H_{int}(t) = -\boldsymbol{V} \cdot E(t)
\end{equation}
where $\boldsymbol{V}$ is the transition dipole operator.  For all of the transitions considered here,
the x-ray wavelength is more than an order of magnitude larger than the core orbital, so the
dipole approximation is justified.

Contributions to these signals come in three varieties, the ground-state bleach (GSB), excited-state emission (ESE), and excited-state absorption (ESA).  In the ESE and ESA contributions, the first two pulses act on different sides of the density matrix creating a core-excited population which is then probed via stimulated emission or absorption of the second pulse pair, respectively.  In the GSB term, the first two pulses act on the same side of the density matrix in an up-down fashion, creating a valence-excited wavepacket which is then probed by the third and fourth pulses. Since the lifetimes of core-excited states are very short (less than 10 fs for the cores considered here\cite{zschornack_handbook_2007}) than those for valence excitations, the ESE and ESA terms can be eliminated by restricting $t_2$ to be longer than the core lifetime.  We assume well-separated pulses.

  The four-wave mixing signals are given by

\begin{equation}\begin{split}\label{eq:ki}
    S_{\ki}&(-\Omega_1, -\Omega_2, \Omega_3) = \\ & \sum_{e,e',g'} \frac{
      \mathcal{E}_1^{*}(\omega_{e' g})
      \mathcal{E}_2(\omega_{e' g'  })
      \mathcal{E}_4^{*}(\omega_{eg'})
      \mathcal{E}_3(\omega_{e g  })
      V_{ge}V_{eg'}V_{g'e'}V_{e'g}
    }{
      ( -\Omega_1 + \omega_{e'g} + i \Gamma_{e'})
      ( -\Omega_2 + \omega_{g' g} + i \Gamma_{g'})
      ( \Omega_3 - \omega_{e g'  } + i \Gamma_{e  })
    }.
\end{split}\end{equation}
where $\mathcal{E}_i(\omega)$ is the spectral envelope of the \thh{i} pulse, $\omega_{rs}$ is the transition frequency between states $r$ and $s$, and $\Gamma_e$ is the inverse lifetime.  The $\kii$ signal is likewise given by
\begin{equation}\begin{split}\label{eq:kii}
    S_{\ki}&(-\Omega_1, -\Omega_2, \Omega_3) = \\ & \sum_{e,e',g'} \frac{
      \mathcal{E}_1^{*}(\omega_{e' g})
      \mathcal{E}_2(\omega_{e' g'  })
      \mathcal{E}_4^{*}(\omega_{eg'})
      \mathcal{E}_3(\omega_{e g  })
      V_{ge}V_{eg'}V_{g'e'}V_{e'g}
    }{
      ( -\Omega_1 + \omega_{e'g} + i \Gamma_{e'})
      ( -\Omega_2 + \omega_{g' g} + i \Gamma_{g'})
      ( \Omega_3 - \omega_{e g'  } + i \Gamma_{e  })
    }.
\end{split}\end{equation}

Four-wave mixing is a natural extension that sheds light on the SXRS process. The two-pulse and three-pulse versions of SXRS are pump-probe experiments wherein the system interacts twice with each pulse.  In the photon echo experiment, the pump pulse is split into two, pulses $\kkk_1$ and $\kkk_2$, each of which interact with the system just once.  Likewise the probe pulse is split into pulses $\kkk_3$ and $\kkk_4$.

In the simulations presented in Sec. \ref{sec:sim}, we perform ensemble averaging over the random distribution of molecular orientations using the tensor formalism.\cite{sma_reference,andrews_three-dimensional_1977}

\section{Two-pulse SXRS Stimulated X-Ray Raman Signals}\label{app:2pSXRS}

The integrated SXRS signal (Fig. \ref{fig:1}) is one dimensional.  The signal is given by\cite{zhang:194306,healion_simulation_2011}
\begin{equation}\begin{split}\label{eq:sxrsfreq2a}
S_{I2P-SXRS}(t_2) &= \Im \int_{-\infty}^{\infty} \mathrm{d}t_3 \mathcal{E}_2^{*}(t_3) P^{(3)}(t_2,t_3) \\
 &=-\sum_{g'}  \left( \frac{ \alpha''_{2;gg'}\alpha_{1;g'g}}{\Omega_2 - \omega_{g'g}+i \Gamma_{g'}} + \frac{ (\alpha''_{2;gg'}\alpha_{1;g'g})^*}{\Omega_2 + \omega_{g'g}+i \Gamma_{g'}}  \right).
\end{split}\end{equation}
where
\begin{equation}\begin{split}
    \label{eq:alphaiso}
{\alpha}_{j} &= {\alpha_j}'+i{\alpha_j}'' \\
    &= \sum_{e,g',g''} \ket{g'}
    \frac{V_{g' e} V_{e g''} }{2\pi}
    \int _{-\infty }^{\infty } d\omega \frac{\mathcal{E}_j^*\left(\omega\right)\mathcal{E}_j\left(\omega+\omega_{g'g''}\right)}{\omega-\omega_{eg'}+i \Gamma_{e}} \bra{g''}
\end{split}\end{equation}
is the effective isotropic polarizability averaged over the spectral envelope of the \thh{j} ultrashort pulse, $\mathcal{E}_j$.  The two terms in Eq. \ref{eq:sxrsfreq2a} have peaks in the positive and negative $\Omega_2$ regions, respectively, which carry the same information.  We thus only plot the positive $\Omega_2$ region, retaining only the first term in Eq. \ref{eq:sxrsfreq2a}.

Note that the signal depends only on the anti-Hermitian part of $\alpha_2$.  This is because it represents the change in the intensity of the probe pulse \emph{integrated over all frequencies} due to the presence of the pump.  If the probe is off resonance, then $\alpha_2'' = 0$ and its net absorption by the system will vanish regardless of the pump.  The transmission of an off-resonant probe is a parametric process, energy can be exchanged between different modes within the probe bandwidth, but will not change its integrated value.

A two-dimensional signal can be obtained with two pulses by sending the probe through a spectrometer, and recording the dispersed spectrum as a function of the interpulse delay, which is then Fourier transformed numerically.  The D2P-SXRS signal is given by
\begin{equation}\label{eq:sxrsD1}
S_{D2P-SXRS}(t_2,\Omega_3) = \Im  \mathcal{E}_2^{*}(\Omega_3) P^{(3)}(t_2,\Omega_3)
\end{equation}
which, after a Fourier transform with respect to the delay time, evaluates to

\begin{equation}\label{eq:fd1}
S_{D2P-SXRS}(\Omega_2,\Omega_3) =  \sum_{e,g'} \frac{i V_{ge}V_{eg'} (\alpha_{1;g'g}) }{\Omega_2-\omega_{g'g}+i \Gamma_{g'}}  \left(\frac{\mathcal{E}_2(\Omega_3+\omega_{gg'})\mathcal{E}_2^{*}(\Omega_3)}{\Omega_3-\omega_{eg}+i \Gamma_{e}}-\frac{\mathcal{E}^{*}_2(\Omega_3+\omega_{g'g})\mathcal{E}_2(\Omega_3)}{\Omega_3-\omega_{eg'}-i \Gamma_{e}} \right).
\end{equation}
It can easily be verified that
\begin{equation}
S_{I2P-SXRS}(\Omega_2)=\int \mathrm{d}\Omega_3 S_{D2P-SXRS}(\Omega_2,\Omega_3).
\end{equation}
Eq. \ref{eq:fd1} shows that while the frequency-dispersed signal becomes weaker when the probe is off resonance, it does not vanish.  However, the integrated signal does vanish, as is shown in Appendix \ref{app:alpha}.

\section{The Effective Polarizability}
\label{app:alpha}
The effective polarizability is defined as
\begin{equation}
    \label{eq:alpha}
    \alpha_{j;g' g}= \sum_e
    \frac{V_{g' e}V_{e g}}{2\pi}
    \int _{-\infty }^{\infty } d\omega \frac{\mathcal{E}_j^*\left(\omega\right)\mathcal{E}_j\left(\omega+\omega_{g'g}\right)}{\omega-\omega_{eg'}+i \Gamma_{e}}
\end{equation}
This operator is non-Hermitian, and we may write the matrix elements of its conjugate transpose as
\begin{equation}\begin{split}
    \label{eq:alphadagger}
    \alpha^\dagger_{j;g' g} =&  (\alpha_{j;g g'})^*\\
    =& \sum_e
    \frac{V_{g' e}V_{e g}}{2\pi}
    \int _{-\infty }^{\infty } d\omega \frac{\mathcal{E}_j\left(\omega\right)\mathcal{E}^*_j\left(\omega+\omega_{gg'}\right)}{\omega-\omega_{eg}-i \Gamma_{e}} \\
    =& \sum_e
    \frac{V_{g' e}V_{e g}}{2\pi}
    \int _{-\infty }^{\infty } d\omega \frac{\mathcal{E}^*_j\left(\omega\right)\mathcal{E}_j\left(\omega+\omega_{g'g}\right)}{\omega-\omega_{eg'}-i \Gamma_{e}}
\end{split}\end{equation}
It is useful to explicitly split $\alpha_j$ into Hermitian and anti-Hermitian parts as
\begin{equation}\begin{split}\label{eq:alphaH_AH}
\alpha'_j =&(\alpha_j+\alpha^\dagger_j)/2 \\
i\alpha''_j = &(\alpha_j-\alpha^\dagger_j)/2
\end{split} .\end{equation}
whose matrix elements are
\begin{equation}\begin{split}
    \label{eq:alphaH}
    \alpha'_{j;g' g}&= \sum_e
    \frac{V_{g' e}V_{e g}}{4\pi}
    \int _{-\infty }^{\infty } d\omega \mathcal{E}_j^*\left(\omega\right)\mathcal{E}_j\left(\omega+\omega_{g'g}\right) \left(\frac{1}{\omega-\omega_{eg'}+i \Gamma_{e}}+\frac{1}{\omega-\omega_{eg'}-i \Gamma_{e}}\right) \\
    &= \sum_e
    \frac{V_{g' e}V_{e g}}{2\pi}
    \int _{-\infty }^{\infty } d\omega \mathcal{E}_j^*\left(\omega\right)\mathcal{E}_j\left(\omega+\omega_{g'g}\right)\frac{ \omega-\omega_{eg}}{\Gamma_{e}^2+(\omega-\omega_{eg'})^2}
\end{split},\end{equation}
and
\begin{equation}\begin{split}
    \label{eq:alphaAH}
    i\alpha''_{j;g' g}&= \sum_e
    \frac{V_{g' e}V_{e g}}{4\pi}
    \int _{-\infty }^{\infty } d\omega \mathcal{E}_j^*\left(\omega\right)\mathcal{E}_j\left(\omega+\omega_{g'g}\right) \left(\frac{1}{\omega-\omega_{eg'}+i \Gamma_{e}}-\frac{1}{\omega-\omega_{eg'}-i \Gamma_{e}}\right) \\
    &= \sum_e
    \frac{ V_{g' e}V_{e g}}{2\pi}
    \int _{-\infty }^{\infty } d\omega \mathcal{E}_j^*\left(\omega\right)\mathcal{E}_j\left(\omega+\omega_{g'g}\right)\frac{ -i \Gamma_{e}}{\Gamma_{e}^2+(\omega-\omega_{eg'})^2}
\end{split}.\end{equation}
In the off-resonant case, where $\mathcal{E}_j(\omega_{eg'})=0$, the Hermitian part of $\alpha_j$ does not go to zero while the anti-Hermitian part does.  This can be seen explicitly by taking considering Eqs. \ref{eq:alphaH} and \ref{eq:alphaAH} in the limit where $\Gamma_{e} \rightarrow 0$.

We can also easily show that the D2P-SXRS signal vanishes when the probe is off resonance,
\begin{equation}\begin{split}
S_{SXRS}(\Omega_2)= \sum_{e,g'} & \frac{i V_{ge}V_{eg'} (\alpha_{1;g'g}) }{\Omega_2-\omega_{g'g}+i \Gamma_{g'}} \\ & \times \left(\int \mathrm{d}\Omega_3\frac{\mathcal{E}_2(\Omega_3+\omega_{gg'})\mathcal{E}_2^{*}(\Omega_3)}{\Omega_3-\omega_{eg}+i \Gamma_{e}}-\int \mathrm{d}\Omega_3\frac{\mathcal{E}^{*}_2(\Omega_3+\omega_{g'g})\mathcal{E}_2(\Omega_3)}{\Omega_3-\omega_{eg'}-i \Gamma_{e}} \right) \end{split}\end{equation}
First we make the substitution in the second integral $\Omega_3' = \Omega_3+\omega_{g'g}$, giving
\begin{equation}
S_{SXRS}(\Omega_2)= \sum_{e,g'}  \frac{i V_{ge}V_{eg'} (\alpha_{1;g'g}) }{\Omega_2-\omega_{g'g}+i \Gamma_{g'}}  \left(\int \mathrm{d}\Omega_3\frac{\mathcal{E}_2(\Omega_3+\omega_{gg'})\mathcal{E}_2^{*}(\Omega_3)}{\Omega_3-\omega_{eg}+i \Gamma_{e}}-\int \mathrm{d}\Omega_3'\frac{\mathcal{E}^{*}_2(\Omega_3')\mathcal{E}_2(\Omega_3'+\omega_{gg'})}{\Omega_3'-\omega_{eg}-i \Gamma_{e}} \right), \end{equation}
we then combine the two integrals, renaming $\Omega_3'$ to $\Omega_3$, giving
\begin{equation}\begin{split}
S_{SXRS}(\Omega_2)= \sum_{e,g'} & \frac{i V_{ge}V_{eg'} (\alpha_{1;g'g}) }{\Omega_2-\omega_{g'g}+i \Gamma_{g'}}  \int \mathrm{d}\Omega_3 \mathcal{E}_2(\Omega_3+\omega_{gg'})\mathcal{E}_2^{*}(\Omega_3) \left(\frac{1}{\Omega_3-\omega_{eg}+i \Gamma_{e}}-\frac{1}{\Omega_3'-\omega_{eg}-i \Gamma_{e}} \right) \\
= \sum_{e,g'} & \frac{ V_{ge}V_{eg'} (\alpha_{1;g'g}) }{\Omega_2-\omega_{g'g}+i \Gamma_{g'}}  \int \mathrm{d}\Omega_3 \mathcal{E}_2(\Omega_3+\omega_{gg'})\mathcal{E}_2^{*}(\Omega_3) \frac{2 \Gamma_e}{(\Omega_3-\omega_{eg})^2+\Gamma_e^2}
\end{split}\end{equation}
The fraction in the integrand will be non-negligible only near resonance, but if the probe-pulse spectral envelope vanishes near resonance, the integral will vanish as well.

\section{Three-pulse SXRS Stimulated X-Ray Raman Signals}\label{app:3pSXRS}
  The I3P-SXRS signal (Fig. \ref{fig:2}), written in a consistent notation to that above, is
\begin{equation}\label{eq:2dsxrs1}
S_{I3P-SXRS}(\Omega_2,\Omega_4) =  \sum_{g',g''} \frac{ i \alpha_{1;g'g} }{\Omega_2-\omega_{g'g}+i \Gamma_{g'}} \left(\frac{\alpha_{2;gg''}^\dagger \alpha''_{3;g''g'}}{\Omega_4-\omega_{g'g''}+i \Gamma_{g'}} -\frac{\alpha''_{3;gg''} \alpha_{2;g''g'}}{\Omega_4-\omega_{g''g}+i \Gamma_{g''}} \right).
\end{equation}
Eq. \ref{eq:2dsxrs1} contains only positive $\Omega_2$ resonances.  Note the symmetry, $S_{I3P-SXRS}(-\Omega_2,-\Omega_4)=\left(S_{I3P-SXRS}(\Omega_2,\Omega_4)\right)^*$.

To obtain the expression for the frequency-dispersed three-pulse SXRS (D3P-SXRS) signal, we simply need to ``unpack'' the probe polarizability in Eq. \ref{eq:2dsxrs1}, just as Eq. \ref{eq:fd1} was obtained from Eq. \ref{eq:sxrsfreq2a}.
\begin{equation}\begin{split} \label{eq:2dsxrs1b}
& S_{D3P-SXRS}(\Omega_2,\Omega_4,\Omega_5) =  \sum_{e,g',g''} \\ & \qquad \frac{ i \alpha_{1;g'g} }{\Omega_2-\omega_{g'g}+i \Gamma_{g'}} \frac{\alpha_{2;g,g''}^\dagger }{\Omega_4-\omega_{g'g''}+i \Gamma_{g'}} \left(\frac{\mathcal{E}_3(\Omega_5+\omega_{g''g'})\mathcal{E}_3^{*}(\Omega_5)}{\Omega_5-\omega_{eg''}+i \Gamma_{e}}-\frac{\mathcal{E}^{*}_3(\Omega_5+\omega_{g'g''})\mathcal{E}_3(\Omega_5)}{\Omega_5-\omega_{eg'}-i \Gamma_{e}} \right) \\
& \qquad-\frac{ i \alpha_{1;g'g} }{\Omega_2-\omega_{g'g}+i \Gamma_{g'}}\frac{ \alpha_{2;g'',g'}}{\Omega_4-\omega_{g''g}+i \Gamma_{g''}} \left(\frac{\mathcal{E}_3(\Omega_5+\omega_{gg''})\mathcal{E}_3^{*}(\Omega_5)}{\Omega_5-\omega_{eg}+i \Gamma_{e}}-\frac{\mathcal{E}^{*}_3(\Omega_5+\omega_{g''g})\mathcal{E}_3(\Omega_5)}{\Omega_5-\omega_{eg''}-i \Gamma_{e}} \right)
\end{split}\end{equation}
\end{widetext}

\end{document}